\documentclass[%
reprint,
superscriptaddress,
amsmath,amssymb,
aps,
]{revtex4-2}

\usepackage{graphicx}
\usepackage{dcolumn}
\usepackage{bm}
\usepackage{siunitx}
\usepackage[separate-uncertainty=true]{siunitx}
\usepackage[inter-unit-product=\ensuremath{{\cdot}}]{siunitx}
\usepackage{comment}
\usepackage{xcolor}
\usepackage[french]{babel}

\begin{document}

\preprint{APS/123-QED}

\title{Nonmonotonic Radiative Heat Transfer in the Transition from Far Field to Near Field}

\author{Victor Guillemot}
 \affiliation{Institut Langevin, ESPCI Paris, Université PSL , CNRS, 75005 Paris, France}

\author{Riccardo Messina}%
 \affiliation{Laboratoire Charles Fabry, UMR 8501, Institut d'Optique, CNRS, Universit\'{e} Paris-Saclay, 2 Avenue Augustin Fresnel, 91127 Palaiseau Cedex, France.}

\author{Valentina Krachmalnicoff}%
 \affiliation{Institut Langevin, ESPCI Paris, Université PSL , CNRS, 75005 Paris, France}

\author{Rémi Carminati}%
 \affiliation{Institut Langevin, ESPCI Paris, Université PSL , CNRS, 75005 Paris, France}
  \affiliation{Institut d’Optique Graduate School, Paris-Saclay University, 91127 Palaiseau, France}

\author{Philippe Ben-Abdallah}%
 \affiliation{Laboratoire Charles Fabry, UMR 8501, Institut d'Optique, CNRS, Universit\'{e} Paris-Saclay, 2 Avenue Augustin Fresnel, 91127 Palaiseau Cedex, France.}

\author{Wilfrid Poirier}%
    \email{wilfrid.poirier@lne.fr}
 \affiliation{Laboratoire national de métrologie et d'essais (LNE), 29 avenue Roger Hennequin, 78197 Trappes, France}

\author{Yannick De Wilde}%
\email{yannick.dewilde@espci.fr}
 \affiliation{Institut Langevin, ESPCI Paris, Université PSL , CNRS, 75005 Paris, France}

             
\begin{abstract}
We present high precision measurements of the radiative heat transfer of a glass microsphere immersed in a thermal bath in vacuum facing three different planar substrates (SiO$_2$, SiC and Au), which exhibit very different optical behaviors in the infrared region. Using a thermoresistive probe on a cantilever, we show the nonmonotonic behavior of the radiative flux between the microsphere and its environment when the microsphere is brought closer to the substrate in the far-field to near-field transition regime. We demonstrate that this unexpected behavior is related to the singularities of dressed emission mechanisms in this three-body system sphere-substrate-bath with respect to the separation distance. 

\end{abstract}

\maketitle


When two bodies at different temperatures come close together in vacuum, the radiative heat transfer (RHT) between them can become orders of magnitude larger than that predicted by classical radiative transfer based on Planck’s law, due to the contribution of evanescent waves such as surface polaritons (SPs) \cite{mulet_enhanced_2002,joulain_surface_2005}. The deviation from classical radiative heat transfer prediction occurs when the gap between the surfaces gets smaller than the thermal wavelength set by Wien's law ($\lambda_\mathrm{Wien}(\SI{300}{K}) \approx\SI{10}{\micro \meter}$). The presence of SP modes enables energy to tunnel through the gap between the surfaces, which results in an increase of the radiated heat flux as distance shortens. The significant increase of RHT in the near field strongly depends on the optical properties of the materials and on the related spectral matching of the local density of states \cite{joulain_definition_2003} above each interface in the infrared domain. Since the work of Polder and Van Hove in 1971 \cite{polder_theory_1971}, a lot of experimental works  have demonstrated the contribution of surface modes to RHT, in different geometries including sphere-plane \cite{rousseau_radiative_2009,narayanaswamy_near-field_2008} or plane/plane \cite{domoto_experimental_1970,bernardi_radiative_2016, fiorino_giant_2018}. This effect has been probed with different sensors, such as mechanical flux sensors \cite{rousseau_radiative_2009} or electrical ones \cite{lucchesi_near-field_2021,kittel_near-field_2005}. The study of small-scale heat transfer proved to be of crucial importance for thermal photovoltaics \cite{mittapally_near-field_2021,lucchesi_near-field_2021} or thermal surface probing \cite{reihani_quantitative_2022}. 
\begin{figure}[!h]
    \centering
    \includegraphics[width = 0.4\textwidth]{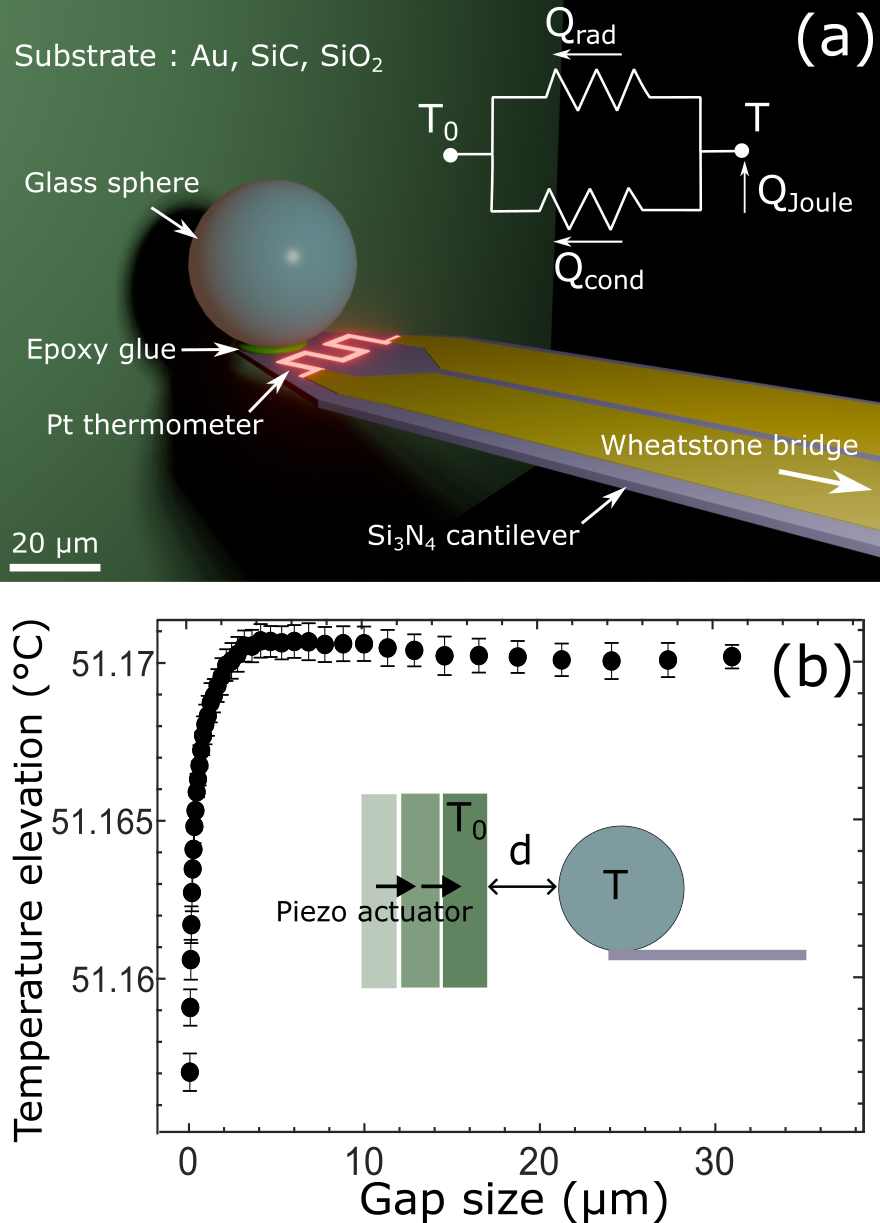}
    \caption{\textbf{Schematic representation of the setup. }(a) A \SI{42.3 \pm 1.1}{\micro \meter} diameter glass microsphere faces a flat substrate made of gold (Au), silicon carbide (SiC) or fused silica (SiO$_2$). The sphere is glued at the end of a \SI{200}{\micro \meter} long cantilever, with a platinum serpentine thermometer. A Joule power $Q_\mathrm{Joule}$ is produced by means of an AC electrical current circulating through the serpentine. The variation of resistance of the Pt thermometer is measured by a metrological Wheatstone bridge. The inset shows the thermal equivalent circuit. (b) Temperature measurement of the sphere $\Delta T = T - T_0$ as a function of the gap size $d$, for a silica substrate where $T_0$ is fixed at 25°C.}
    \label{fig:1}
\end{figure}

Numerous experimental demonstrations of theoretical predictions have focused on submicron gaps, where near-field effects largely dominate. In this work, we study thermal transfer phenomena in a sphere-plane geometry at gap sizes $d$ ranging from \SI{50}{\micro \meter} (i.e. a distance comparable to the size of the microsphere) to contact. Covering distances from the far field to the near field enables us to highlight the peculiarities of the transition between two radiative regimes, which are respectively driven by propagating and evanescent waves. Silicon carbide (SiC), silica (SiO$_2$) and gold (Au), which are materials with very different optical behavior in the infrared, are considered as planar substrate.  

As sketched in Fig.~\ref{fig:1}(a), a \SI{42.3 \pm 1.1}{\micro \meter} diameter borosilicate microsphere is glued on a custom-designed probe consisting of a \SI{200}{\micro \meter} long silicon nitride cantilever (Kelvin Nanotechnologies) equipped with a platinum thermometer (see section 1 in supplementary). The cantilever is horizontal and the sphere, at temperature $T$, faces the planar substrate, which is held vertically. Its temperature is fixed and precisely controlled by a  Peltier element stabilized at the temperature of the surrounding bath, $T_0 = $ \SI{25}{\celsius}. The position of the microsphere with respect to the plane is set by a piezoelectric nanopositioner whose axis of motion is normal to the plane and which has a resolution of \SI{5}{\nano \meter}. All measurements are performed under high vacuum condition ($P = $ \SI{e-6}{\milli \bar}) to avoid any conductive or convective heat transfer between the sphere and the plane. Mechanical vibrations are reduced by placing the setup on a hydraulic pressurized table.\\ 
The thermometer, of electrical resistance $R$, consists of a thin serpentine layer of platinum patterned near the extremity of the cantilever ($R =$ \SI{293}{\ohm} at $T_0$ = \SI{25}{\celsius}). It is used both to apply a net temperature difference $\Delta T = T - T_0$ between the sphere and the plane by Joule heating and to measure precisely the temperature of the probe thanks to the temperature coefficient of resistance of Platinum. We use a metrological Wheatstone bridge \cite{doumouro_quantitative_2021} to precisely measure $R$ with a sub-$\mathrm{m \Omega}$ precision. The Platinum serpentine is fed with an alternating current at \SI{10}{\kilo \hertz}  which is above the thermal frequency cutoff of the probe. As a result, the sphere temperature does not oscillate and depends on the average Joule power. The unbalanced voltage of the bridge, measured with a lock-in detection, is directly proportional to the platinum serpentine temperature variation (see section 2 in supplementary).   
The inset in Fig.~\ref{fig:1}(a) describes the thermal equivalent circuit of the sphere in interaction with the substrate and the cantilever, assuming that the sphere is at the temperature of the serpentine platinum strip $ T > T_0$. The radiative flux $Q_\mathrm{rad}(d)$ between the sphere and its full surrounding environment is given by:
\begin{equation}
    Q_\mathrm{rad}(d) = Q_\mathrm{Joule}(d) - G_\mathrm{cond} \Delta T(d),
    \label{eq:conductance}
\end{equation}
where $G_\mathrm{cond}$ is the thermal conductance of the cantilever and $Q_\mathrm{Joule}$ is the average Joule heating power injected at the top of the sphere. 
The thermal conductance through the cantilever $G_\mathrm{cond}$ is determined from a linear fit by applying different Joule heating $Q_\mathrm{Joule}$ on the sphere far from the plane and measuring the associated temperature rise. We obtain $G_\mathrm{cond} = $ \SI{5.79(0.01)} {\micro \watt .\kelvin^{-1}} . Note that this calibration also includes the far-field radiative conductance which is orders of magnitude smaller than the cantilever conductance due to conduction. Since the temperature coefficient of the Pt thermometer, separately calibrated in a thermalized oven, is $k=$\SI{1.34}{\kelvin\per\ohm}, the sub-m$\Omega$ precision of the bridge corresponds to a \SI{}{\milli \kelvin} precision in the temperature measurements (see supplementary section 2).
%
\begin{figure}[!h]
    \centering
    \includegraphics[width = 0.44 \textwidth]{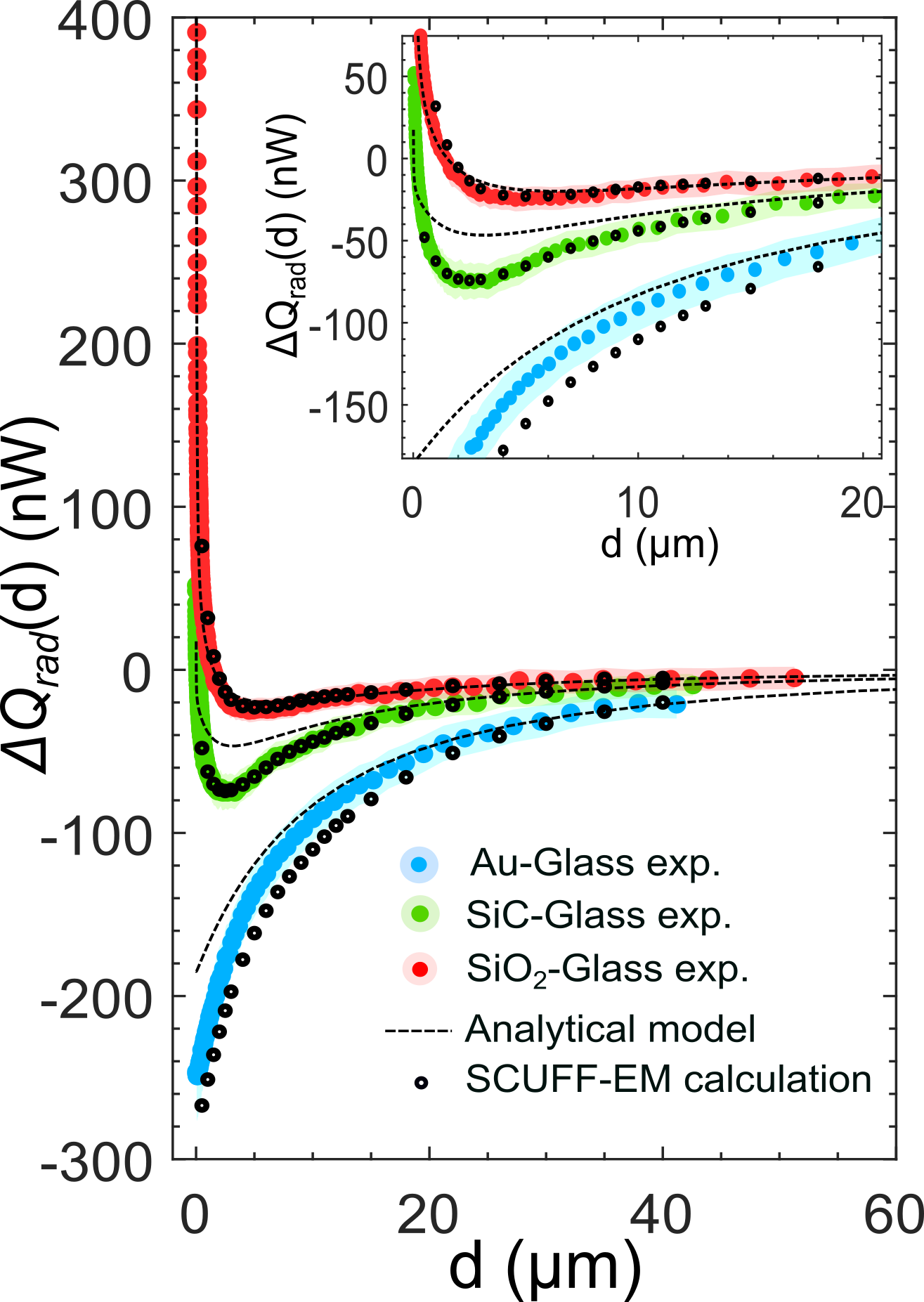}
    \caption{\textbf{Gap-size dependence of the flux variation between a $\bm{42.3}$ $ \bm{\mathrm{\mu m}} $ diameter borosilicate microsphere and three different substrates for $\bm{\Delta T = 70}$ $\bm{\mathrm{ K}}$} (a) Measurements for gold (blue dots), SiC (green dots) and SiO$_2$ (red dots) with uncertainties shown as colored regions. The three measurements exhibit a different behavior because of the competition between two transmission mechanisms. In the case of $\mathrm{SiC}$ and $\mathrm{SiO_2}$ substrates, when the gap size decreases, the flux first decreases and then rises sharply because of SPs resonances.  In the case of a gold substrate the flux decreases over the entire distance range. Simulation results obtained with SCUFF-EM (black dots) are in good agreement with the measurement.  Dashed curves are the result of the theoretical model detailed in the main text. The inset shows an enlargement in the far-to-near field crossover region.} 
    \label{fig:2}
\end{figure}
Figure ~\ref{fig:1}(b) shows an example of the temperature measured over distances ranging from a few tens of micrometers down to contact with an integration time of \SI{1}{\second}. The uncertainty shown has been statistically evaluated using the standard deviation of the mean. 
Because the device is subject to small ambient variation of temperature, we proceed to a ``back-and-forth" approach: for each position we measure the absolute value of the flux $Q_\mathrm{rad}(d)$ at a distance $d$ from the surface, and then, immediately after, the background flux $Q_\mathrm{rad}(d_\mathrm{BG})$ far from the surface. Here, the distance $d_\mathrm{BG}$ is equal to \SI{60}{\micro \meter}, which is the maximal amplitude of the piezo stage. The quantity $\Delta Q_\mathrm{rad}(d) = Q_\mathrm{rad}(d) - Q_\mathrm{rad}(d_\mathrm{BG})$ is therefore the flux variation with distance. This "back-and-forth" technique has two advantages: first, we directly access the distance dependence of the flux by suppressing the background. Second, it allows us to neglect thermal drifts since $Q_\mathrm{rad}(d)$  and $Q_\mathrm{rad}(d_\mathrm{BG})$ are measured within a few seconds for each new position $d$ of the approach curve. We notice that, in our study, the measurement of $\Delta Q_\mathrm{rad}(d)$ is equivalent to the measurement of $ Q_\mathrm{rad}(d)$ to within a constant. 
Figure ~\ref{fig:2} shows the measurement of $\Delta Q_\mathrm{rad}(d)$ for the three substrates over a distance varying from \SI{50}{\micro \meter} to contact with a temperature difference $\Delta T = $ \SI{70}{\kelvin}. Remarkably the flux varies with the distance very differently depending on the substrate. For the SiO$_2$ substrate the flux decreases slowly when the sphere gets closer to the substrate, reaches a minimum around \SI{5}{\micro \meter} and then increases sharply. For the SiC substrate, we observe the same behavior with a stronger decrease and a minimum flux around \SI{3}{\micro \meter}. Finally, we observe a continuous decrease of the flux when the sphere approaches the gold substrate. Other measurements at $\Delta T = $ \SI{20}{\celsius} and $\Delta T = $ \SI{40}{\celsius} exhibit the same behavior (see Fig. S6 in the supplementary).  

\begin{figure}[!h]
    \centering
    \includegraphics[width = 0.45\textwidth]{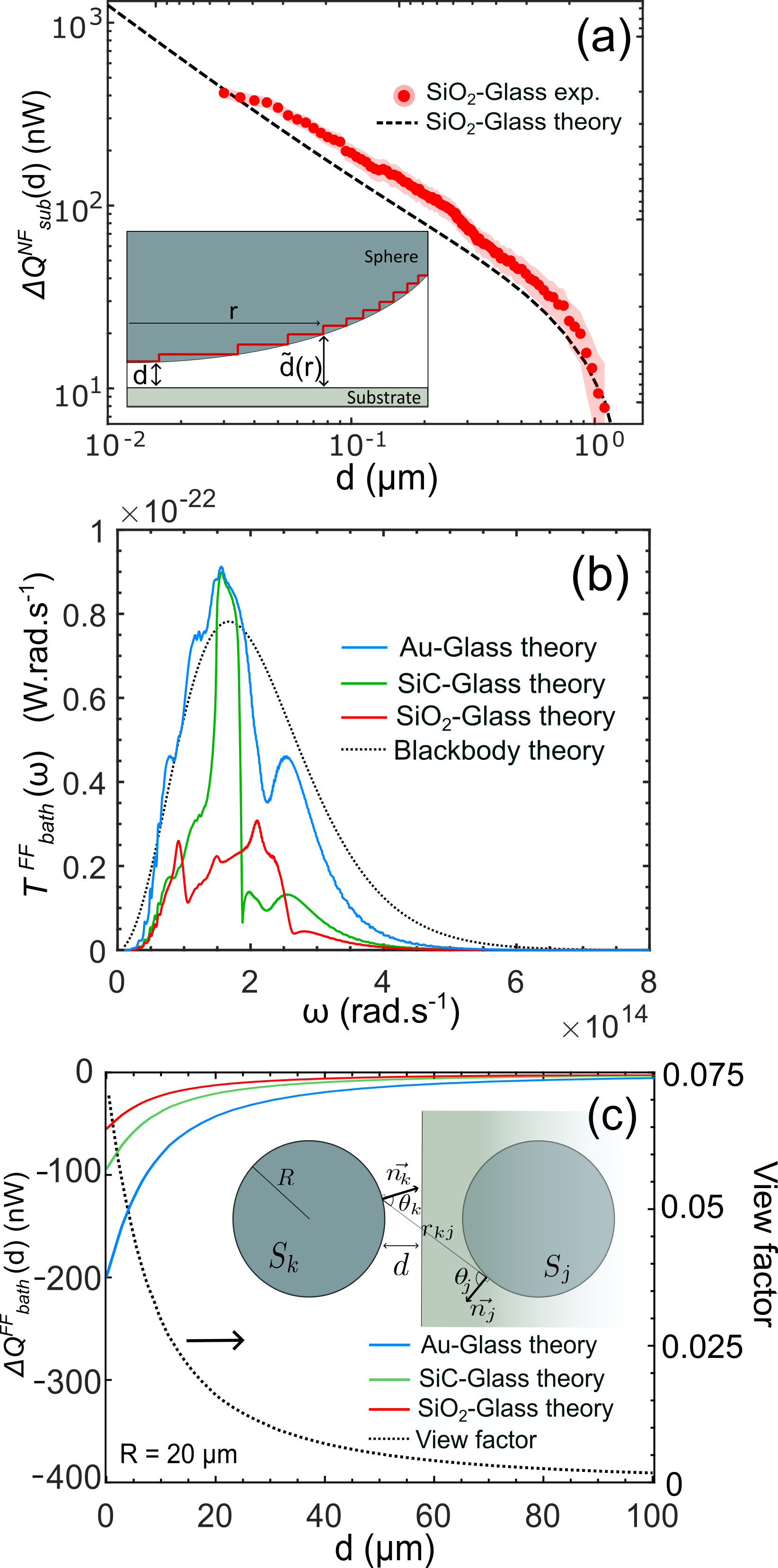}
    \caption{\textbf{Analytical calculation of $\bm{\Delta Q_\mathrm{bath}^\mathrm{FF}(d)}$ and $\bm{\Delta Q_\mathrm{sub}^\mathrm{NF}(d)}$ } (a) Measurement with uncertainty of the near-field flux $\Delta Q_\mathrm{rad}(d)$ between the microsphere and a $\mathrm{SiO_2}$ substrate. The dashed line corresponds to the calculation of $\Delta Q_\mathrm{sub}^\mathrm{NF}(d)$ using Derjaguin approximation shown in the inset. (b) Far-field spectral flux between the sphere and the bath, $\mathcal{T^\mathrm{FF}_\mathrm{bath}}$, for the three different substrates compared to the blackbody spectral flux (dotted curve) at $ d = $ \SI{50}{\micro \meter}. (c) Variation of the far-field flux $ \Delta Q_\mathrm{bath}^\mathrm{FF}(d)$ as a function of distance. The dotted curve corresponds to the value of the view factor between two spheres separated by a distance $2d$ and the inset shows the notations used in Eq.~\eqref{eqn:viewfactor}.}
    \label{fig:3}
\end{figure}

In order to understand the physical phenomena behind the nonmonotonic behavior of $\Delta Q_\mathrm{rad}(d)$ we propose an analytical model where the net radiative heat flux exchanged by the sphere with both the substrate (which is optically smooth) and the surrounding bath considered as a blackbody and composed of the semi-infinite half space facing the substrate, can be written as the sum of three contributions: 
\begin{equation}
    Q_\mathrm{rad}(d) = Q_\mathrm{bath}^\mathrm{FF}(d) + Q_\mathrm{sub}^\mathrm{FF} + Q_\mathrm{sub}^\mathrm{NF}(d) ,
    \label{eqn:Q}
\end{equation}
where $Q_\mathrm{bath}^\mathrm{FF}(d)$, $Q_\mathrm{sub}^\mathrm{FF}$  and $Q_\mathrm{sub}^\mathrm{NF}(d)$ are the radiative heat fluxes exchanged between the microsphere and the surrounding bath in far field, the substrate in far field and the substrate in near field, respectively. We can write this expression in terms of transmission coefficients using Landauer's formalism:
\begin{multline}
     Q_\mathrm{rad}(d) =  \int_{0}^{\infty} \frac{\mathrm{d}\omega }{2\pi}  \Delta \Theta(\omega,T,T_0) \\ \times [\mathcal{T^\mathrm{FF}_\mathrm{bath}} (\omega,d) + \mathcal{T^\mathrm{FF}_\mathrm{sub}} (\omega) + \mathcal{T^\mathrm{NF}_\mathrm{sub}} (\omega,d) ],
     \label{eqn:Q_tau}
\end{multline}
where $\Delta \Theta(\omega,T,T_0) = \Theta(\omega,T) - \Theta(\omega,T_0)$, $\Theta(\omega,T) = \hbar \omega / (e^{\hbar \omega / k_b T} - 1)$ is the mean energy of the Planck oscillator at temperature $T$. $\mathcal{T^\mathrm{FF}_\mathrm{bath}}$, $\mathcal{T^\mathrm{FF}_\mathrm{sub}}$ and $\mathcal{T^\mathrm{NF}_\mathrm{sub}}$ denote the transmission coefficients using the same notation as in Eq.~\eqref{eqn:Q}. We can now express these three transmission coefficients. First, the transmission coefficient in near field can be calculated, in the sphere-plane Derjaguin approximation (valid for $d \ll R $)   \cite{Derjaguin,otey_numerically_2011}, in terms of the plane-plane transmission coefficient $\mathcal{\tilde{T}^\mathrm{NF}_\mathrm{sub}}$ obtained without approximation in the fluctuational-electrodynamics framework \cite{joulain_surface_2005} according to: 
\begin{equation}
    \mathcal{T^\mathrm{NF}_\mathrm{sub}} (\omega, d) = \int_{0}^{R} \mathrm{d} r \ \mathcal{\tilde{T}^\mathrm{NF}_\mathrm{sub}} (\omega, \tilde{d}(r)) 2\pi r ,
\end{equation}
\begin{equation}
    \mathcal{\tilde{T}^\mathrm{NF}_\mathrm{sub}} (\omega, d) = \sum_{s,p} \int_{\omega / c}^{\infty} \frac{\mathrm{d} \kappa}{2 \pi}   \frac{4 \kappa \mathrm{Im}(r^{s,p}_{13}) \mathrm{Im}(r^{s,p}_{23})}{\vert 1 - r^{s,p}_{12} r^{s,p}_{13} e^{- 2 i k_\perp d} \vert ^2} e^{-2 \mathrm{Im}(k_\perp) d },
    \label{eqn:T_nf_sub}
\end{equation}
where, $r^{s,p}_{ij}$ are the Fresnel coefficients of reflection for each polarization $(s,p)$.  The indices 1 and 2 correspond to the sphere and the substrate while index 3 corresponds to vacuum. The quantity $\tilde{d}(r) = d + R - \sqrt{R^2 - r^2}$ is the distance between the substrate and the fraction of the sphere surface assimilated to an infinitesimal ring of radius $r$ in the Derjaguin approximation [see the inset in Fig.~\ref{fig:3}(a)]. 
The two far-field spectral transmission coefficients can be expressed in terms of apparent emissivities and view factor \cite{campbell_radiant-interchange_nodate} using a radiometric approach valid for $d \gg \lambda_\mathrm{Wien}$ and $\delta \ll R $, $\delta$ being the skin depth of electromagnetic radiation in materials. Neglecting multiple reflections between the sphere and the substrate, and assuming a specular reflection at the substrate interface, we find that: 
\begin{equation}
    \mathcal{T^\mathrm{FF}_\mathrm{bath}} (\omega,d) = R^2  \frac{\omega^2}{c^2}\varepsilon_{\mathrm{sphere}}(\omega) \left[ 1 + \rho_{\mathrm{sub}}(\omega) (1 - 2f(2d))\right],
    \label{eqn:T_ff_bath}
\end{equation}
\begin{equation}
    \mathcal{T^\mathrm{FF}_\mathrm{sub}} (\omega) =    R^2 \frac{\omega^2}{c^2}\varepsilon_{\mathrm{sphere}}(\omega) \varepsilon_{\mathrm{sub}}(\omega) ,
    \label{eqn:T_ff_sub}
\end{equation}
\begin{equation}
    f(2d) = \frac{1}{S_k} \int_{S_k} \int_{S_j} \frac{\cos(\theta_k) \cos(\theta_j)}{\pi r_{jk}^2}  \, \mathrm{d}A_j \mathrm{d}A_k,
    \label{eqn:viewfactor}
\end{equation}
with $\varepsilon_\mathrm{sphere}(\omega)$ the spectral emissivity of the silica microsphere, $\rho_{\mathrm{sub}}(\omega)$ the spectral reflectivity of the substrate (see Fig. S8 in supplementary) and $f(2d)$ the view factor between two spheres separated by a distance $2d$ (plotted in dotted lines in Fig.~\ref{fig:3}(c)). The value of $\varepsilon_\mathrm{sphere}(\omega)$ have been calculated with Mie theory \cite{kattawar_radiation_1970} as the efficiency coefficients of absorption of the microsphere (see Fig. S10 in  supplementary). The transmission coefficient $\mathcal{T^\mathrm{FF}_\mathrm{bath}}(d =$ \SI{50}{\micro \meter}) is represented in Fig.~\ref{fig:3}(b) for the three substrates and for a perfect blackbody for comparison. After full-spectrum integration we obtain the far-field flux variation $\Delta Q_\mathrm{bath}^\mathrm{FF}(d) = Q_\mathrm{bath}^\mathrm{FF}(d) - Q_\mathrm{bath}^\mathrm{FF}(d\to\infty)$ in Fig.~\ref{fig:3}(c).  We observe that this flux decreases when the gap size shrinks. We see in Eq.~\eqref{eqn:T_ff_bath} that the reduction of flux exchanged between the sphere with the rest of the environment is proportional to the flux that would be exchanged between two spheres in far field separated by a distance $2d$. The nonmonotonic behavior observed in this work is directly linked to the fact that the microsphere is hotter than the surrounding bath. Previous experiments \cite{rousseau_radiative_2009,narayanaswamy_near-field_2008} could not have detected this distinctive feature, as the hot element in their setup was the substrate, while the sphere remained at the temperature of the surrounding bath. Finally, because the substrate is nearly infinite, the contribution $Q_\mathrm{sub}^\mathrm{FF}$ does not depend on the distance $d$, which means that $\Delta Q_\mathrm{sub}^\mathrm{FF}(d) = Q_\mathrm{sub}^\mathrm{FF}(d) - Q_\mathrm{sub}^\mathrm{FF}(d\to\infty) = 0$. The flux exchanged in far-field between the sphere and the substrate does not influence the variation of radiative flux $\Delta Q_\mathrm{rad}(d)$. The sum of the near-field contribution (Eq. (5)) with the far-field contributions (Eq. (6) and Eq. (7)) successfully captures the non-monotonic behavior of $\Delta Q_\mathrm{rad}(d)$ but fails to quantitatively predict it in the crossover region, where both models describing these contributions exhibit theoretical limitations. The best agreement with our measurements was found when dividing all the near-field contributions $\Delta Q_\mathrm{NF}(d)$ by a factor 2.5, which is the only adjustable parameter in our modeling of the far-field-near-field transition.  Note that a similar scaling of the Derjaguin contribution  has been  reported to fit experimental data obtained in the same geometry \cite{yan_surface_2023,song_enhancement_2015}.
The scaled values of $ \Delta Q_\mathrm{sub}^\mathrm{NF}(d) = Q_\mathrm{sub}^\mathrm{NF}(d) - Q_\mathrm{sub}^\mathrm{NF}(d\to\infty)$ are represented for the $\mathrm{SiO_2}$ substrate within the last micrometer in Fig.~\ref{fig:3}(a) (see Fig. S7 for the two other substrates in supplementary). At such short distances, the calculated curve matches very well the measurements of $\Delta Q_\mathrm{rad}(d)$. We observe that the measurements seem to tend toward the $1/d$ behavior predicted by the Derjaguin approximation \cite{rousseau_radiative_2009}.

The sum of the near-field contribution with the two far-field contributions for the three different planar substrates is represented by the dashed curves in Fig.~\ref{fig:2}. The analytical calculation has been done  down to d = \SI{1}{\nano \meter}. The competition between the geometric flux reduction exchanged with the surrounding bath and the increase of flux exchanged with the substrate due to the near-field coupling leads to the non-monotonic behavior observed experimentally. One can note that in the case of the gold substrate, because of the poor spectral matching of polaritons of gold with those of SiO$_2$, the increase of the flux due to near field is expected to appear around $d = $ \SI{15}{\nano \meter}. This is a distance which is not achievable in our setup due to the microsphere roughness. This is why we only observe the decrease of the flux corresponding to $\Delta Q_\mathrm{bath}^\mathrm{FF}(d)$, since $f(2d)$ increases until $d= 0$. Thanks to the remarkable agreement with experimental data at various values of $\Delta T$ (see Fig. S6 in supplementary), we see that the single-reflection model is sufficient to explain the non-monotonic behavior of the flux with distance. Moreover, exact plane-plane geometry calculations that account for multiple reflections cannot explain the phenomenon since they give a monotonic increase in flux with distance (see Fig. S11 in supplementary). This model clearly accounts for the influence of the immediate environment of the sphere on its far-field thermal emission, both toward the substrate and the surrounding thermal bath,  highlighting the major role played by the dressing of a thermal emitter in the radiative exchanges with other bodies in its vicinity.

However, our calculations rely on two approximations : the radiometric approach using geometrical optics, which is only valid for  $d > \lambda$, $\delta \ll R $ and  the Derjaguin approximation when $d \ll R $ \cite{otey_numerically_2011}. 
Because the model is not fully quantitative, a more rigorous calculation of transmission coefficients in the sphere-plane geometry has been performed at the cost of a significant numerical effort, taking into account the wave character of electromagnetic fields. The black dots reported in Fig.~\ref{fig:2} are the result of a numerical calculation performed with SCUFF-EM, a free open-source software based on the boundary element method for the description of scattering of electromagnetic waves (i.e. without either of the two aforementioned approximations)\cite{rodriguez_fluctuating-surface-current_2013,SCUFF1,SCUFF2} . In the simulation, a \SI{40}{\micro \meter}-diameter sphere is placed in vacuum in front of a planar substrate (see Fig. S12 in supplementary). We computed the heat flux exchanged by the sphere for each radiation frequency and a given set of gap sizes. After integration over the full frequency spectrum we obtained very good agreement between calculation and measurements for the three different materials. The nonmonotonic behavior of $\Delta Q_\mathrm{rad}(d)$ is clearly visible for SiC and SiO$_2$. Note that the SCUFF-EM calculation gives the net value of the flux, while our measurement gives the variation of flux with the distance after subtraction of the large distance background flux. For this reason, we have shifted the SCUFF-EM curves vertically so that they coincide with our measurements. The dielectric functions used in this paper are shown on Fig. S8 in the supplementary materials. Because the borosilicate sphere is made  of almost 80\% silica, the dielectric function of $\mathrm{SiO_2}$ was used to describe the electromagnetic properties of the glass microsphere.

In summary, precise measurements of temperature variation of a glass microsphere above a planar substrate using a thermoresistive local probe allowed one to quantify the variation of flux exchanged with its surrounding  environment. As a result, we have demonstrated that the thermal emissivity of an object is significantly altered by the presence of a nearby substrate, which in turn modifies its radiative exchanges with the surrounding environment. This modification leads to a nonmonotonic variation in these exchanges during the transition between the near-field and far-field regimes, affecting the thermalization process at such separation distances. This effect could enable precise tuning and control of radiative heat transfer and thermalization dynamics, with important implications for thermal management and energy conversion.
\\
\\
\textit{Aknowledgements} - This work was supported by the ”Investissements d’Avenir” program launched by the French Government (Labex WiFi) and by the Agence Nationale de la Recherche (NBODHEAT Project No. ANR-21-CE30-0030). The authors are grateful to Patricia Al Alam (Inst. Langevin) for sharing experimental advice on SThM measurements.

\bibliography{Biblio.bib}

\begin{thebibliography}{23}%
\makeatletter
\providecommand \@ifxundefined [1]{%
 \@ifx{#1\undefined}
}%
\providecommand \@ifnum [1]{%
 \ifnum #1\expandafter \@firstoftwo
 \else \expandafter \@secondoftwo
 \fi
}%
\providecommand \@ifx [1]{%
 \ifx #1\expandafter \@firstoftwo
 \else \expandafter \@secondoftwo
 \fi
}%
\providecommand \natexlab [1]{#1}%
\providecommand \enquote  [1]{``#1''}%
\providecommand \bibnamefont  [1]{#1}%
\providecommand \bibfnamefont [1]{#1}%
\providecommand \citenamefont [1]{#1}%
\providecommand \href@noop [0]{\@secondoftwo}%
\providecommand \href [0]{\begingroup \@sanitize@url \@href}%
\providecommand \@href[1]{\@@startlink{#1}\@@href}%
\providecommand \@@href[1]{\endgroup#1\@@endlink}%
\providecommand \@sanitize@url [0]{\catcode `\\12\catcode `\$12\catcode `\&12\catcode `\#12\catcode `\^12\catcode `\_12\catcode `\%12\relax}%
\providecommand \@@startlink[1]{}%
\providecommand \@@endlink[0]{}%
\providecommand \url  [0]{\begingroup\@sanitize@url \@url }%
\providecommand \@url [1]{\endgroup\@href {#1}{\urlprefix }}%
\providecommand \urlprefix  [0]{URL }%
\providecommand \Eprint [0]{\href }%
\providecommand \doibase [0]{https://doi.org/}%
\providecommand \selectlanguage [0]{\@gobble}%
\providecommand \bibinfo  [0]{\@secondoftwo}%
\providecommand \bibfield  [0]{\@secondoftwo}%
\providecommand \translation [1]{[#1]}%
\providecommand \BibitemOpen [0]{}%
\providecommand \bibitemStop [0]{}%
\providecommand \bibitemNoStop [0]{.\EOS\space}%
\providecommand \EOS [0]{\spacefactor3000\relax}%
\providecommand \BibitemShut  [1]{\csname bibitem#1\endcsname}%
\let\auto@bib@innerbib\@empty
\bibitem [{\citenamefont {Mulet}\ \emph {et~al.}(2002)\citenamefont {Mulet}, \citenamefont {Joulain}, \citenamefont {Carminati},\ and\ \citenamefont {Greffet}}]{mulet_enhanced_2002}%
  \BibitemOpen
  \bibfield  {author} {\bibinfo {author} {\bibfnamefont {J.-P.}\ \bibnamefont {Mulet}}, \bibinfo {author} {\bibfnamefont {K.}~\bibnamefont {Joulain}}, \bibinfo {author} {\bibfnamefont {R.}~\bibnamefont {Carminati}},\ and\ \bibinfo {author} {\bibfnamefont {J.-J.}\ \bibnamefont {Greffet}},\ }\bibfield  {title} {\bibinfo {title} {Enhanced radiative heat transfer at nanometric distances},\ }\href {https://doi.org/10.1080/10893950290053321} {\bibfield  {journal} {\bibinfo  {journal} {Microscale Thermophysical Engineering}\ }\textbf {\bibinfo {volume} {6}},\ \bibinfo {pages} {209} (\bibinfo {year} {2002})}\BibitemShut {NoStop}%
\bibitem [{\citenamefont {Joulain}\ \emph {et~al.}(2005)\citenamefont {Joulain}, \citenamefont {Mulet}, \citenamefont {Marquier}, \citenamefont {Carminati},\ and\ \citenamefont {Greffet}}]{joulain_surface_2005}%
  \BibitemOpen
  \bibfield  {author} {\bibinfo {author} {\bibfnamefont {K.}~\bibnamefont {Joulain}}, \bibinfo {author} {\bibfnamefont {J.-P.}\ \bibnamefont {Mulet}}, \bibinfo {author} {\bibfnamefont {F.}~\bibnamefont {Marquier}}, \bibinfo {author} {\bibfnamefont {R.}~\bibnamefont {Carminati}},\ and\ \bibinfo {author} {\bibfnamefont {J.-J.}\ \bibnamefont {Greffet}},\ }\bibfield  {title} {\bibinfo {title} {Surface electromagnetic waves thermally excited: {Radiative} heat transfer, coherence properties and {Casimir} forces revisited in the near field},\ }\href {https://doi.org/10.1016/j.surfrep.2004.12.002} {\bibfield  {journal} {\bibinfo  {journal} {Surface Science Reports}\ }\textbf {\bibinfo {volume} {57}},\ \bibinfo {pages} {59} (\bibinfo {year} {2005})}\BibitemShut {NoStop}%
\bibitem [{\citenamefont {Joulain}\ \emph {et~al.}(2003)\citenamefont {Joulain}, \citenamefont {Carminati}, \citenamefont {Mulet},\ and\ \citenamefont {Greffet}}]{joulain_definition_2003}%
  \BibitemOpen
  \bibfield  {author} {\bibinfo {author} {\bibfnamefont {K.}~\bibnamefont {Joulain}}, \bibinfo {author} {\bibfnamefont {R.}~\bibnamefont {Carminati}}, \bibinfo {author} {\bibfnamefont {J.-P.}\ \bibnamefont {Mulet}},\ and\ \bibinfo {author} {\bibfnamefont {J.-J.}\ \bibnamefont {Greffet}},\ }\bibfield  {title} {\bibinfo {title} {Definition and measurement of the local density of electromagnetic states close to an interface},\ }\href {https://doi.org/10.1103/PhysRevB.68.245405} {\bibfield  {journal} {\bibinfo  {journal} {Physical Review B}\ }\textbf {\bibinfo {volume} {68}},\ \bibinfo {pages} {245405} (\bibinfo {year} {2003})}\BibitemShut {NoStop}%
\bibitem [{\citenamefont {Polder}\ and\ \citenamefont {Van~Hove}(1971)}]{polder_theory_1971}%
  \BibitemOpen
  \bibfield  {author} {\bibinfo {author} {\bibfnamefont {D.}~\bibnamefont {Polder}}\ and\ \bibinfo {author} {\bibfnamefont {M.}~\bibnamefont {Van~Hove}},\ }\bibfield  {title} {\bibinfo {title} {Theory of {Radiative} {Heat} {Transfer} between {Closely} {Spaced} {Bodies}},\ }\href {https://doi.org/10.1103/PhysRevB.4.3303} {\bibfield  {journal} {\bibinfo  {journal} {Physical Review B}\ }\textbf {\bibinfo {volume} {4}},\ \bibinfo {pages} {3303} (\bibinfo {year} {1971})}\BibitemShut {NoStop}%
\bibitem [{\citenamefont {Rousseau}\ \emph {et~al.}(2009)\citenamefont {Rousseau}, \citenamefont {Siria}, \citenamefont {Jourdan}, \citenamefont {Volz}, \citenamefont {Comin}, \citenamefont {Chevrier},\ and\ \citenamefont {Greffet}}]{rousseau_radiative_2009}%
  \BibitemOpen
  \bibfield  {author} {\bibinfo {author} {\bibfnamefont {E.}~\bibnamefont {Rousseau}}, \bibinfo {author} {\bibfnamefont {A.}~\bibnamefont {Siria}}, \bibinfo {author} {\bibfnamefont {G.}~\bibnamefont {Jourdan}}, \bibinfo {author} {\bibfnamefont {S.}~\bibnamefont {Volz}}, \bibinfo {author} {\bibfnamefont {F.}~\bibnamefont {Comin}}, \bibinfo {author} {\bibfnamefont {J.}~\bibnamefont {Chevrier}},\ and\ \bibinfo {author} {\bibfnamefont {J.-J.}\ \bibnamefont {Greffet}},\ }\bibfield  {title} {\bibinfo {title} {Radiative heat transfer at the nanoscale},\ }\href {https://doi.org/10.1038/nphoton.2009.144} {\bibfield  {journal} {\bibinfo  {journal} {Nature Photonics}\ }\textbf {\bibinfo {volume} {3}},\ \bibinfo {pages} {514} (\bibinfo {year} {2009})}\BibitemShut {NoStop}%
\bibitem [{\citenamefont {Narayanaswamy}\ \emph {et~al.}(2008)\citenamefont {Narayanaswamy}, \citenamefont {Shen},\ and\ \citenamefont {Chen}}]{narayanaswamy_near-field_2008}%
  \BibitemOpen
  \bibfield  {author} {\bibinfo {author} {\bibfnamefont {A.}~\bibnamefont {Narayanaswamy}}, \bibinfo {author} {\bibfnamefont {S.}~\bibnamefont {Shen}},\ and\ \bibinfo {author} {\bibfnamefont {G.}~\bibnamefont {Chen}},\ }\bibfield  {title} {\bibinfo {title} {Near-field radiative heat transfer between a sphere and a substrate},\ }\href {https://doi.org/10.1103/PhysRevB.78.115303} {\bibfield  {journal} {\bibinfo  {journal} {Physical Review B}\ }\textbf {\bibinfo {volume} {78}},\ \bibinfo {pages} {115303} (\bibinfo {year} {2008})}\BibitemShut {NoStop}%
\bibitem [{\citenamefont {Domoto}\ \emph {et~al.}(1970)\citenamefont {Domoto}, \citenamefont {Boehm},\ and\ \citenamefont {Tien}}]{domoto_experimental_1970}%
  \BibitemOpen
  \bibfield  {author} {\bibinfo {author} {\bibfnamefont {G.~A.}\ \bibnamefont {Domoto}}, \bibinfo {author} {\bibfnamefont {R.~F.}\ \bibnamefont {Boehm}},\ and\ \bibinfo {author} {\bibfnamefont {C.~L.}\ \bibnamefont {Tien}},\ }\bibfield  {title} {\bibinfo {title} {Experimental {Investigation} of {Radiative} {Transfer} {Between} {Metallic} {Surfaces} at {Cryogenic} {Temperatures}},\ }\href {https://doi.org/10.1115/1.3449677} {\bibfield  {journal} {\bibinfo  {journal} {Journal of Heat Transfer}\ }\textbf {\bibinfo {volume} {92}},\ \bibinfo {pages} {412} (\bibinfo {year} {1970})}\BibitemShut {NoStop}%
\bibitem [{\citenamefont {Bernardi}\ \emph {et~al.}(2016)\citenamefont {Bernardi}, \citenamefont {Milovich},\ and\ \citenamefont {Francoeur}}]{bernardi_radiative_2016}%
  \BibitemOpen
  \bibfield  {author} {\bibinfo {author} {\bibfnamefont {M.~P.}\ \bibnamefont {Bernardi}}, \bibinfo {author} {\bibfnamefont {D.}~\bibnamefont {Milovich}},\ and\ \bibinfo {author} {\bibfnamefont {M.}~\bibnamefont {Francoeur}},\ }\bibfield  {title} {\bibinfo {title} {Radiative heat transfer exceeding the blackbody limit between macroscale planar surfaces separated by a nanosize vacuum gap},\ }\href {https://doi.org/10.1038/ncomms12900} {\bibfield  {journal} {\bibinfo  {journal} {Nature Communications}\ }\textbf {\bibinfo {volume} {7}},\ \bibinfo {pages} {12900} (\bibinfo {year} {2016})}\BibitemShut {NoStop}%
\bibitem [{\citenamefont {Fiorino}\ \emph {et~al.}(2018)\citenamefont {Fiorino}, \citenamefont {Thompson}, \citenamefont {Zhu}, \citenamefont {Song}, \citenamefont {Reddy},\ and\ \citenamefont {Meyhofer}}]{fiorino_giant_2018}%
  \BibitemOpen
  \bibfield  {author} {\bibinfo {author} {\bibfnamefont {A.}~\bibnamefont {Fiorino}}, \bibinfo {author} {\bibfnamefont {D.}~\bibnamefont {Thompson}}, \bibinfo {author} {\bibfnamefont {L.}~\bibnamefont {Zhu}}, \bibinfo {author} {\bibfnamefont {B.}~\bibnamefont {Song}}, \bibinfo {author} {\bibfnamefont {P.}~\bibnamefont {Reddy}},\ and\ \bibinfo {author} {\bibfnamefont {E.}~\bibnamefont {Meyhofer}},\ }\bibfield  {title} {\bibinfo {title} {Giant {Enhancement} in {Radiative} {Heat} {Transfer} in {Sub}-30 nm {Gaps} of {Plane} {Parallel} {Surfaces}},\ }\href {https://doi.org/10.1021/acs.nanolett.8b00846} {\bibfield  {journal} {\bibinfo  {journal} {Nano Letters}\ }\textbf {\bibinfo {volume} {18}},\ \bibinfo {pages} {3711} (\bibinfo {year} {2018})}\BibitemShut {NoStop}%
\bibitem [{\citenamefont {Lucchesi}\ \emph {et~al.}(2021)\citenamefont {Lucchesi}, \citenamefont {Cakiroglu}, \citenamefont {Perez}, \citenamefont {Taliercio}, \citenamefont {Tournié}, \citenamefont {Chapuis},\ and\ \citenamefont {Vaillon}}]{lucchesi_near-field_2021}%
  \BibitemOpen
  \bibfield  {author} {\bibinfo {author} {\bibfnamefont {C.}~\bibnamefont {Lucchesi}}, \bibinfo {author} {\bibfnamefont {D.}~\bibnamefont {Cakiroglu}}, \bibinfo {author} {\bibfnamefont {J.-P.}\ \bibnamefont {Perez}}, \bibinfo {author} {\bibfnamefont {T.}~\bibnamefont {Taliercio}}, \bibinfo {author} {\bibfnamefont {E.}~\bibnamefont {Tournié}}, \bibinfo {author} {\bibfnamefont {P.-O.}\ \bibnamefont {Chapuis}},\ and\ \bibinfo {author} {\bibfnamefont {R.}~\bibnamefont {Vaillon}},\ }\bibfield  {title} {\bibinfo {title} {Near-{Field} {Thermophotovoltaic} {Conversion} with {High} {Electrical} {Power} {Density} and {Cell} {Efficiency} above 14\%},\ }\href {https://doi.org/10.1021/acs.nanolett.0c04847} {\bibfield  {journal} {\bibinfo  {journal} {Nano Letters}\ }\textbf {\bibinfo {volume} {21}},\ \bibinfo {pages} {4524} (\bibinfo {year} {2021})}\BibitemShut {NoStop}%
\bibitem [{\citenamefont {Kittel}\ \emph {et~al.}(2005)\citenamefont {Kittel}, \citenamefont {Müller-Hirsch}, \citenamefont {Parisi}, \citenamefont {Biehs}, \citenamefont {Reddig},\ and\ \citenamefont {Holthaus}}]{kittel_near-field_2005}%
  \BibitemOpen
  \bibfield  {author} {\bibinfo {author} {\bibfnamefont {A.}~\bibnamefont {Kittel}}, \bibinfo {author} {\bibfnamefont {W.}~\bibnamefont {Müller-Hirsch}}, \bibinfo {author} {\bibfnamefont {J.}~\bibnamefont {Parisi}}, \bibinfo {author} {\bibfnamefont {S.-A.}\ \bibnamefont {Biehs}}, \bibinfo {author} {\bibfnamefont {D.}~\bibnamefont {Reddig}},\ and\ \bibinfo {author} {\bibfnamefont {M.}~\bibnamefont {Holthaus}},\ }\bibfield  {title} {\bibinfo {title} {Near-{Field} {Heat} {Transfer} in a {Scanning} {Thermal} {Microscope}},\ }\href {https://doi.org/10.1103/PhysRevLett.95.224301} {\bibfield  {journal} {\bibinfo  {journal} {Physical Review Letters}\ }\textbf {\bibinfo {volume} {95}},\ \bibinfo {pages} {224301} (\bibinfo {year} {2005})}\BibitemShut {NoStop}%
\bibitem [{\citenamefont {Mittapally}\ \emph {et~al.}(2021)\citenamefont {Mittapally}, \citenamefont {Lee}, \citenamefont {Zhu}, \citenamefont {Reihani}, \citenamefont {Lim}, \citenamefont {Fan}, \citenamefont {Forrest}, \citenamefont {Reddy},\ and\ \citenamefont {Meyhofer}}]{mittapally_near-field_2021}%
  \BibitemOpen
  \bibfield  {author} {\bibinfo {author} {\bibfnamefont {R.}~\bibnamefont {Mittapally}}, \bibinfo {author} {\bibfnamefont {B.}~\bibnamefont {Lee}}, \bibinfo {author} {\bibfnamefont {L.}~\bibnamefont {Zhu}}, \bibinfo {author} {\bibfnamefont {A.}~\bibnamefont {Reihani}}, \bibinfo {author} {\bibfnamefont {J.~W.}\ \bibnamefont {Lim}}, \bibinfo {author} {\bibfnamefont {D.}~\bibnamefont {Fan}}, \bibinfo {author} {\bibfnamefont {S.~R.}\ \bibnamefont {Forrest}}, \bibinfo {author} {\bibfnamefont {P.}~\bibnamefont {Reddy}},\ and\ \bibinfo {author} {\bibfnamefont {E.}~\bibnamefont {Meyhofer}},\ }\bibfield  {title} {\bibinfo {title} {Near-field thermophotovoltaics for efficient heat to electricity conversion at high power density},\ }\href {https://doi.org/10.1038/s41467-021-24587-7} {\bibfield  {journal} {\bibinfo  {journal} {Nature Communications}\ }\textbf {\bibinfo {volume} {12}},\ \bibinfo {pages} {4364} (\bibinfo {year} {2021})}\BibitemShut {NoStop}%
\bibitem [{\citenamefont {Reihani}\ \emph {et~al.}(2022)\citenamefont {Reihani}, \citenamefont {Luan}, \citenamefont {Yan}, \citenamefont {Lim}, \citenamefont {Meyhofer},\ and\ \citenamefont {Reddy}}]{reihani_quantitative_2022}%
  \BibitemOpen
  \bibfield  {author} {\bibinfo {author} {\bibfnamefont {A.}~\bibnamefont {Reihani}}, \bibinfo {author} {\bibfnamefont {Y.}~\bibnamefont {Luan}}, \bibinfo {author} {\bibfnamefont {S.}~\bibnamefont {Yan}}, \bibinfo {author} {\bibfnamefont {J.~W.}\ \bibnamefont {Lim}}, \bibinfo {author} {\bibfnamefont {E.}~\bibnamefont {Meyhofer}},\ and\ \bibinfo {author} {\bibfnamefont {P.}~\bibnamefont {Reddy}},\ }\bibfield  {title} {\bibinfo {title} {Quantitative {Mapping} of {Unmodulated} {Temperature} {Fields} with {Nanometer} {Resolution}},\ }\href {https://doi.org/10.1021/acsnano.1c08513} {\bibfield  {journal} {\bibinfo  {journal} {ACS Nano}\ }\textbf {\bibinfo {volume} {16}},\ \bibinfo {pages} {939} (\bibinfo {year} {2022})}\BibitemShut {NoStop}%
\bibitem [{\citenamefont {Doumouro}\ \emph {et~al.}(2021)\citenamefont {Doumouro}, \citenamefont {Perros}, \citenamefont {Dodu}, \citenamefont {Rahbany}, \citenamefont {Leprat}, \citenamefont {Krachmalnicoff}, \citenamefont {Carminati}, \citenamefont {Poirier},\ and\ \citenamefont {De~Wilde}}]{doumouro_quantitative_2021}%
  \BibitemOpen
  \bibfield  {author} {\bibinfo {author} {\bibfnamefont {J.}~\bibnamefont {Doumouro}}, \bibinfo {author} {\bibfnamefont {E.}~\bibnamefont {Perros}}, \bibinfo {author} {\bibfnamefont {A.}~\bibnamefont {Dodu}}, \bibinfo {author} {\bibfnamefont {N.}~\bibnamefont {Rahbany}}, \bibinfo {author} {\bibfnamefont {D.}~\bibnamefont {Leprat}}, \bibinfo {author} {\bibfnamefont {V.}~\bibnamefont {Krachmalnicoff}}, \bibinfo {author} {\bibfnamefont {R.}~\bibnamefont {Carminati}}, \bibinfo {author} {\bibfnamefont {W.}~\bibnamefont {Poirier}},\ and\ \bibinfo {author} {\bibfnamefont {Y.}~\bibnamefont {De~Wilde}},\ }\bibfield  {title} {\bibinfo {title} {Quantitative {Measurement} of the {Thermal} {Contact} {Resistance} between a {Glass} {Microsphere} and a {Plate}},\ }\href {https://doi.org/10.1103/PhysRevApplied.15.014063} {\bibfield  {journal} {\bibinfo  {journal} {Physical Review Applied}\ }\textbf {\bibinfo {volume} {15}},\ \bibinfo {pages} {014063} (\bibinfo {year} {2021})}\BibitemShut {NoStop}%
\bibitem [{\citenamefont {Derjaguin}\ \emph {et~al.}(1956)\citenamefont {Derjaguin}, \citenamefont {Abrikosova},\ and\ \citenamefont {Lifshitz}}]{Derjaguin}%
  \BibitemOpen
  \bibfield  {author} {\bibinfo {author} {\bibfnamefont {B.~V.}\ \bibnamefont {Derjaguin}}, \bibinfo {author} {\bibfnamefont {I.~I.}\ \bibnamefont {Abrikosova}},\ and\ \bibinfo {author} {\bibfnamefont {E.~M.}\ \bibnamefont {Lifshitz}},\ }\bibfield  {title} {\bibinfo {title} {Direct measurement of molecular attraction between solids separated by a narrow gap.},\ }\href@noop {} {\bibfield  {journal} {\bibinfo  {journal} {Quart. Rev. Chem. Soc.}\ }\textbf {\bibinfo {volume} {10}},\ \bibinfo {pages} {295–329} (\bibinfo {year} {1956})}\BibitemShut {NoStop}%
\bibitem [{\citenamefont {Otey}\ and\ \citenamefont {Fan}(2011)}]{otey_numerically_2011}%
  \BibitemOpen
  \bibfield  {author} {\bibinfo {author} {\bibfnamefont {C.}~\bibnamefont {Otey}}\ and\ \bibinfo {author} {\bibfnamefont {S.}~\bibnamefont {Fan}},\ }\bibfield  {title} {\bibinfo {title} {Numerically exact calculation of electromagnetic heat transfer between a dielectric sphere and plate},\ }\href {https://doi.org/10.1103/PhysRevB.84.245431} {\bibfield  {journal} {\bibinfo  {journal} {Physical Review B}\ }\textbf {\bibinfo {volume} {84}},\ \bibinfo {pages} {245431} (\bibinfo {year} {2011})}\BibitemShut {NoStop}%
\bibitem [{\citenamefont {Campbell}\ and\ \citenamefont {McConnell}(1968)}]{campbell_radiant-interchange_nodate}%
  \BibitemOpen
  \bibfield  {author} {\bibinfo {author} {\bibfnamefont {J.~P.}\ \bibnamefont {Campbell}}\ and\ \bibinfo {author} {\bibfnamefont {D.~G.}\ \bibnamefont {McConnell}},\ }\bibfield  {title} {\bibinfo {title} {Radiant-interchange configuration factors for spherical and conical surfaces to spheres},\ }\href@noop {} {\bibfield  {journal} {\bibinfo  {journal} {Nasa Technical Notes}\ } (\bibinfo {year} {1968})}\BibitemShut {NoStop}%
\bibitem [{\citenamefont {Kattawar}\ and\ \citenamefont {Eisner}(1970)}]{kattawar_radiation_1970}%
  \BibitemOpen
  \bibfield  {author} {\bibinfo {author} {\bibfnamefont {G.~W.}\ \bibnamefont {Kattawar}}\ and\ \bibinfo {author} {\bibfnamefont {M.}~\bibnamefont {Eisner}},\ }\bibfield  {title} {\bibinfo {title} {Radiation from a homogeneous isothermal sphere},\ }\href {https://doi.org/10.1364/AO.9.002685} {\bibfield  {journal} {\bibinfo  {journal} {Applied Optics}\ }\textbf {\bibinfo {volume} {9}},\ \bibinfo {pages} {2685} (\bibinfo {year} {1970})}\BibitemShut {NoStop}%
\bibitem [{\citenamefont {Yan}\ \emph {et~al.}(2023)\citenamefont {Yan}, \citenamefont {Luan}, \citenamefont {Lim}, \citenamefont {Mittapally}, \citenamefont {Reihani}, \citenamefont {Wang}, \citenamefont {Tsurimaki}, \citenamefont {Fan}, \citenamefont {Reddy},\ and\ \citenamefont {Meyhofer}}]{yan_surface_2023}%
  \BibitemOpen
  \bibfield  {author} {\bibinfo {author} {\bibfnamefont {S.}~\bibnamefont {Yan}}, \bibinfo {author} {\bibfnamefont {Y.}~\bibnamefont {Luan}}, \bibinfo {author} {\bibfnamefont {J.~W.}\ \bibnamefont {Lim}}, \bibinfo {author} {\bibfnamefont {R.}~\bibnamefont {Mittapally}}, \bibinfo {author} {\bibfnamefont {A.}~\bibnamefont {Reihani}}, \bibinfo {author} {\bibfnamefont {Z.}~\bibnamefont {Wang}}, \bibinfo {author} {\bibfnamefont {Y.}~\bibnamefont {Tsurimaki}}, \bibinfo {author} {\bibfnamefont {S.}~\bibnamefont {Fan}}, \bibinfo {author} {\bibfnamefont {P.}~\bibnamefont {Reddy}},\ and\ \bibinfo {author} {\bibfnamefont {E.}~\bibnamefont {Meyhofer}},\ }\bibfield  {title} {\bibinfo {title} {Surface {Phonon} {Polariton}-{Mediated} {Near}-{Field} {Radiative} {Heat} {Transfer} at {Cryogenic} {Temperatures}},\ }\href {https://doi.org/10.1103/PhysRevLett.131.196302} {\bibfield  {journal} {\bibinfo  {journal} {Physical Review Letters}\ }\textbf {\bibinfo {volume} {131}},\ \bibinfo {pages} {196302} (\bibinfo {year}
  {2023})}\BibitemShut {NoStop}%
\bibitem [{\citenamefont {Song}\ \emph {et~al.}(2015)\citenamefont {Song}, \citenamefont {Ganjeh}, \citenamefont {Sadat}, \citenamefont {Thompson}, \citenamefont {Fiorino}, \citenamefont {Fernández-Hurtado}, \citenamefont {Feist}, \citenamefont {Garcia-Vidal}, \citenamefont {Cuevas}, \citenamefont {Reddy},\ and\ \citenamefont {Meyhofer}}]{song_enhancement_2015}%
  \BibitemOpen
  \bibfield  {author} {\bibinfo {author} {\bibfnamefont {B.}~\bibnamefont {Song}}, \bibinfo {author} {\bibfnamefont {Y.}~\bibnamefont {Ganjeh}}, \bibinfo {author} {\bibfnamefont {S.}~\bibnamefont {Sadat}}, \bibinfo {author} {\bibfnamefont {D.}~\bibnamefont {Thompson}}, \bibinfo {author} {\bibfnamefont {A.}~\bibnamefont {Fiorino}}, \bibinfo {author} {\bibfnamefont {V.}~\bibnamefont {Fernández-Hurtado}}, \bibinfo {author} {\bibfnamefont {J.}~\bibnamefont {Feist}}, \bibinfo {author} {\bibfnamefont {F.~J.}\ \bibnamefont {Garcia-Vidal}}, \bibinfo {author} {\bibfnamefont {J.~C.}\ \bibnamefont {Cuevas}}, \bibinfo {author} {\bibfnamefont {P.}~\bibnamefont {Reddy}},\ and\ \bibinfo {author} {\bibfnamefont {E.}~\bibnamefont {Meyhofer}},\ }\bibfield  {title} {\bibinfo {title} {Enhancement of near-field radiative heat transfer using polar dielectric thin films},\ }\href {https://doi.org/10.1038/nnano.2015.6} {\bibfield  {journal} {\bibinfo  {journal} {Nature Nanotechnology}\ }\textbf {\bibinfo {volume} {10}},\ \bibinfo
  {pages} {253} (\bibinfo {year} {2015})}\BibitemShut {NoStop}%
\bibitem [{\citenamefont {Rodriguez}\ \emph {et~al.}(2013)\citenamefont {Rodriguez}, \citenamefont {Reid},\ and\ \citenamefont {Johnson}}]{rodriguez_fluctuating-surface-current_2013}%
  \BibitemOpen
  \bibfield  {author} {\bibinfo {author} {\bibfnamefont {A.~W.}\ \bibnamefont {Rodriguez}}, \bibinfo {author} {\bibfnamefont {M.~T.~H.}\ \bibnamefont {Reid}},\ and\ \bibinfo {author} {\bibfnamefont {S.~G.}\ \bibnamefont {Johnson}},\ }\bibfield  {title} {\bibinfo {title} {Fluctuating-surface-current formulation of radiative heat transfer: {Theory} and applications},\ }\href {https://doi.org/10.1103/PhysRevB.88.054305} {\bibfield  {journal} {\bibinfo  {journal} {Physical Review B}\ }\textbf {\bibinfo {volume} {88}},\ \bibinfo {pages} {054305} (\bibinfo {year} {2013})}\BibitemShut {NoStop}%
\bibitem [{\citenamefont {{Homer Reid}}\ and\ \citenamefont {{Johnson}}(2013)}]{SCUFF1}%
  \BibitemOpen
  \bibfield  {author} {\bibinfo {author} {\bibfnamefont {M.~T.}\ \bibnamefont {{Homer Reid}}}\ and\ \bibinfo {author} {\bibfnamefont {S.~G.}\ \bibnamefont {{Johnson}}},\ }\bibfield  {title} {\bibinfo {title} {{Efficient Computation of Power, Force, and Torque in BEM Scattering Calculations}},\ }\href@noop {} {\bibfield  {journal} {\bibinfo  {journal} {ArXiv e-prints}\ } (\bibinfo {year} {2013})},\ \Eprint {https://arxiv.org/abs/1307.2966} {arXiv:1307.2966 [physics.comp-ph]} \BibitemShut {NoStop}%
\bibitem [{\citenamefont {"\texttt{http://github.com/homerreid/scuff-EM}}()}]{SCUFF2}%
  \BibitemOpen
  \bibfield  {author} {\bibinfo {author} {\bibnamefont {"\texttt{http://github.com/homerreid/scuff-EM}}},\ }\href@noop {} {\ }\BibitemShut {NoStop}%
\end{thebibliography}%

\end{document}


\maketitle

\clearpage

\section{Experimental setup}

\begin{figure}[!h]
    \centering
    \includegraphics[width=0.5\linewidth]{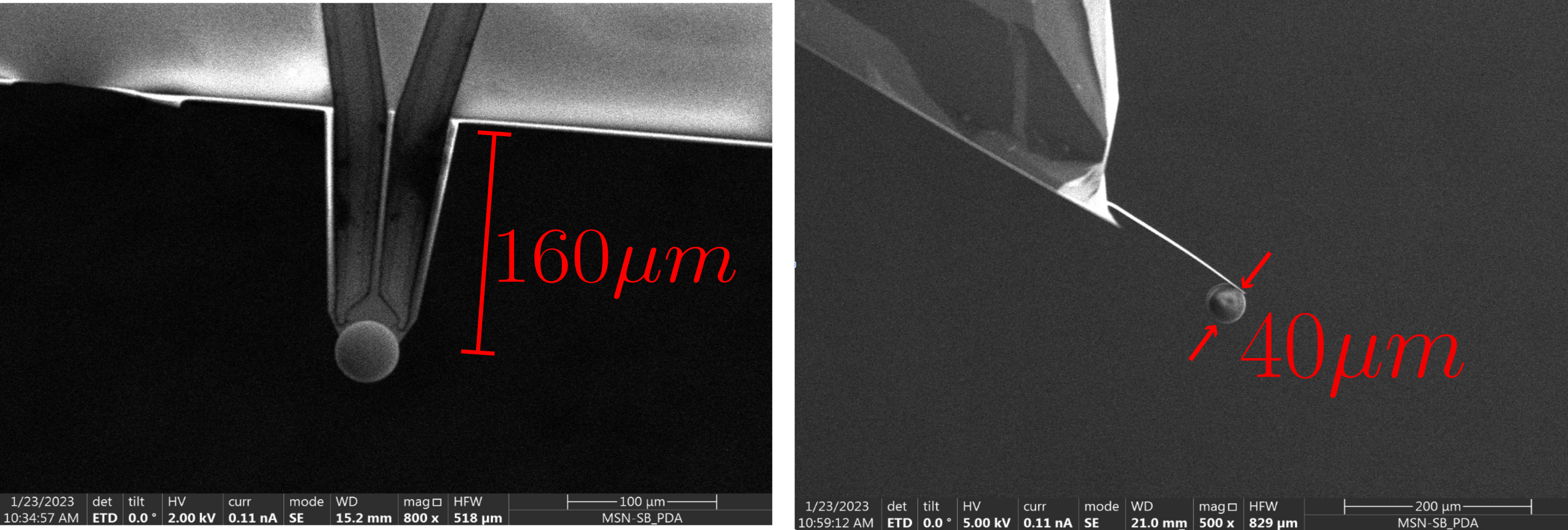}
    \caption{SEM images of the cantilever with the glued \SI{40}{\micro \meter } sphere }
    \label{fig:MEB}
\end{figure}

The scanning thermal probe is made of a Si$_3$N$_4$ (Kelvin Nanotechnologies) cantilever. 
The microsphere we used is a \SI{40}{\micro \meter} diameter NIST$^\mathrm{TM}$ standard borosillicate sphere. In order to reduce the roughness of the sphere we proceed to ultrasonic washing in isopropanol before gluing the bead to the cantilever using thermally resistant epoxy glue. SEM observation of the washed beads gives us the roughness which is typically tens of nanometers.  
Figure S.~\ref{fig:MEB} shows SEM images of the cantilever from top and side view.

\begin{figure}[!h]
    \centering
    \includegraphics[width=0.7\linewidth]{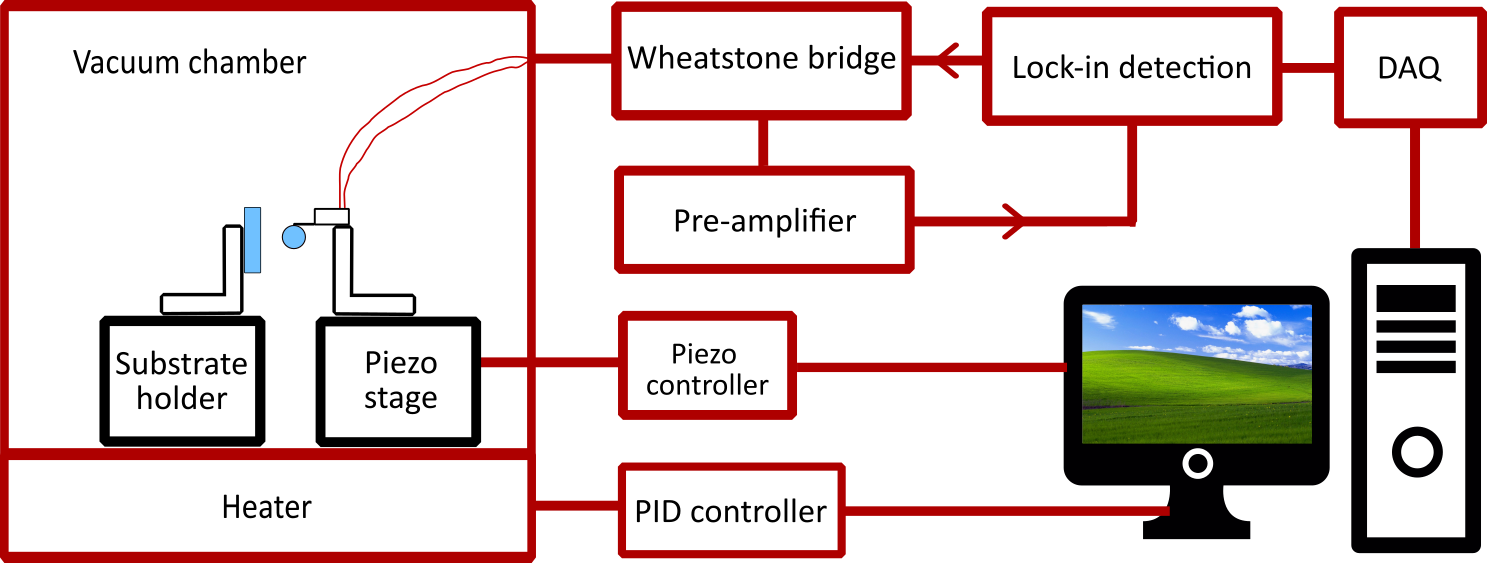}
    \caption{Schematic of the experimental setup}
    \label{fig:setup}
\end{figure}

A schematic of the setup is represented on Fig.~\ref{fig:setup}. The cantilever and the substrate are placed in a thermalized vacuum chamber. The position of the sphere with respect to the plane is controlled by a PiezoJena stage.

\section{Temperature measurement}

\begin{figure}[!h]
    \centering
    \includegraphics[width=0.8\linewidth]{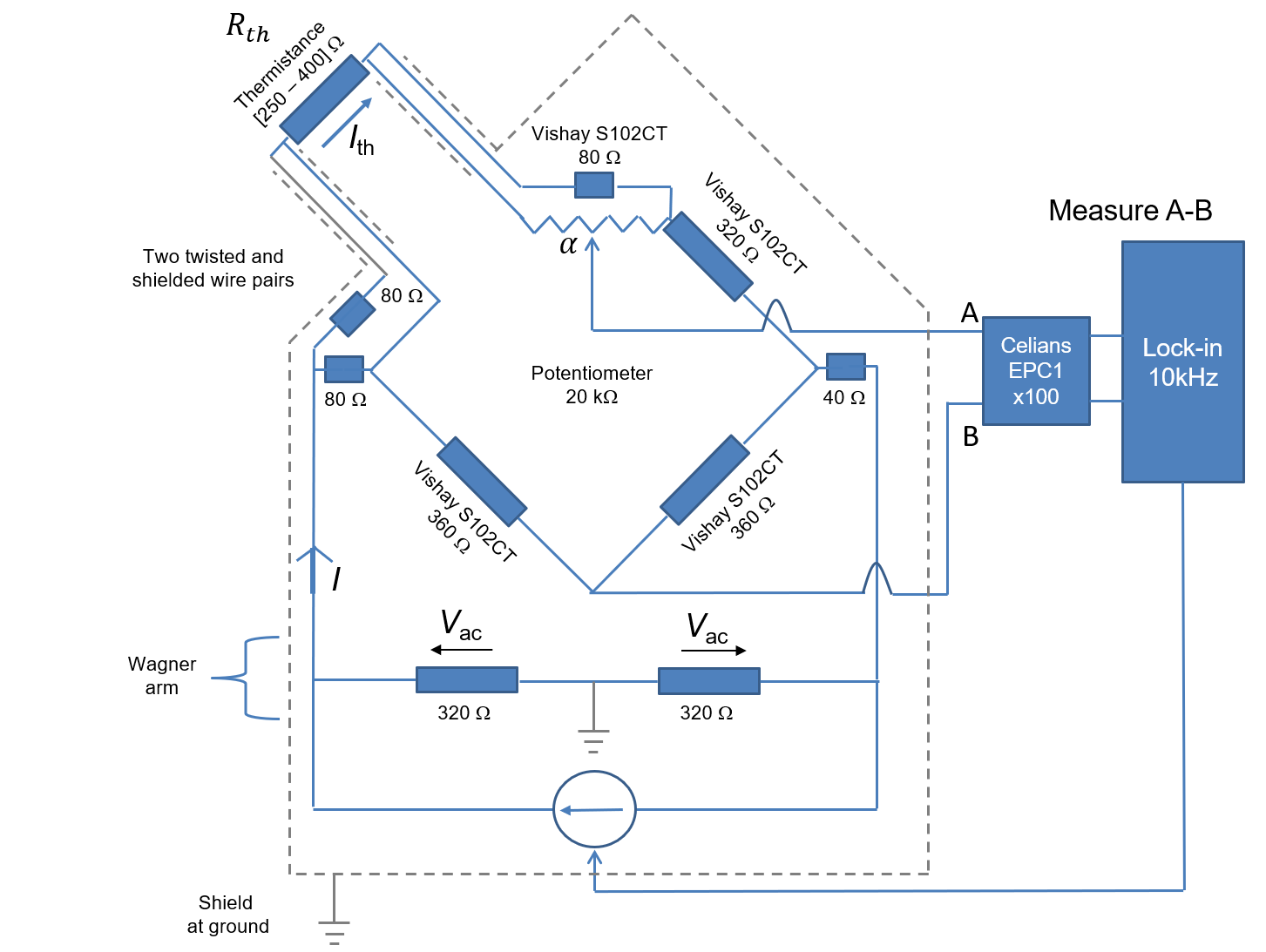}
    \caption{Schematic representation of the Wheatstone bridge}
    \label{fig:pont}
\end{figure}

We use a Wheatstone bridge (see Fig.~\ref{fig:pont}) to measure the resistance value, $R$, of the thermometer, which is obtained from the three other reference resistors and the voltage unbalance of the bridge. The setting $\alpha^\mathrm{eq}$ of the \SI{20}{\kilo \ohm} potentiometer in parallel with the \SI{80}{\ohm} resistor is used to balance the bridge ($V_\mathrm{AB} \approx 0$) in order to accommodate for thermometer resistance value away from that of reference resistors. Detection of a low-voltage signal allows choosing the highest sensitivity range of the voltage detector. The Wheatstone bridge being equipped with a Wagner arm with a midpoint at ground permits both $V_\mathrm{A}$ and $V_\mathrm{B}$ detection voltages to be close to the ground potential. This allows the absence of a common-mode voltage, which is favorable for having better sensitivity and accuracy. First, the bridge is made of S102CT Vishay resistors having a relative drift below $10^{-5}$/year and a low-temperature coefficient below 0.6 ppm/K. The electronic circuits delivering the biasing voltage are based on precision operational amplifiers supplied by stabilized voltage sources powered by rechargeable batteries. Second, the thermometer is connected with four wires in a way (a triangle of two \SI{80}{\ohm} resistors on one side and one \SI{80}{\ohm} resistor and the potentiometer on the other side) that makes the bridge balance more immune to any change (for example, caused by temperature) of the wire resistances. Finally, pairs of high and low potential wires are independently twisted and placed in a shield at ground as the case of the bridge instrument. Because of the midpoint at ground of the Wagner arm, current leakages between wires due to capacitance and insulating resistance are redirected towards the ground without changing the balance of the bridge.
The Wheatstone bridge is biased with an AC voltage signal delivered by the oscillator of a lock-in detector. The output unbalance voltage of the bridge is amplified by a CELIANS EPC1 low-noise amplifier with a gain 100, then measured by the lock-in detector and finally collected by the computer with a National Instrument DAQ. The resistance value $R$ of the thermometer is given by the relationship : 
\begin{equation}
    R = \frac{360 (400 - 160 \alpha^\mathrm{eq}) - 377600 \gamma}{360 + 800 \gamma},\label{eq1}
\end{equation}
where $\gamma = V_\mathrm{AB}/V_\mathrm{AC}$, $V_\mathrm{AB}$ is the amplitude of the ac unbalance voltage measured by the lock-in detector, $V_\mathrm{AC}$ is the AC voltage of frequency $f_\mathrm{AC}\simeq10$ kHz applied to the Wheatstone bridge, and $\alpha^\mathrm{eq} \in [0, ..., 1]$ is the potentiometer setting. One therefore obtains:
\begin{equation}
    R = 400 - 160 \alpha^\mathrm{eq} \label{eq2},
\end{equation}
where $\alpha^\mathrm{eq}$ is the equilibrium setting of the potentiometer ensuring the bridge balance ($\gamma=0$). The current $I_\mathrm{AC}$ that circulates through the thermal sensor is given by :
\begin{equation}
    I_\mathrm{AC} = \frac{720 V_\mathrm{AC}}{377600 + 800 R}\label{eq3}
\end{equation}
Relationships \ref{eq1}, \ref{eq2} and \ref{eq3} are also those obtained under DC current polarization of the bridge. They are valid under AC current polarization if the thermometer has negligible resistance oscillation due to self heating. This happens in case of small AC current or of large AC current at a frequency above the thermal cutoff frequency of the thermometer. In other case, the resistance of thermometer has a significant phase shifted resistance component oscillating at frequency $2f_\mathrm{AC}$.  The circulation of the current $I_\mathrm{AC}$ through this $2f_\mathrm{AC}$ component leads to two voltage contributions, one at frequency $f_\mathrm{AC}$ and one at frequency $3f_\mathrm{AC}$. It results that the setting of the potentiometer, $\alpha^\mathrm{eq}$, which ensures bridge balance ($V_\mathrm{AB}=0$) at fundamental frequency depends on $f_\mathrm{AC}$.  

\begin{figure}[!h]
    \centering
    \includegraphics[width=0.9\linewidth]{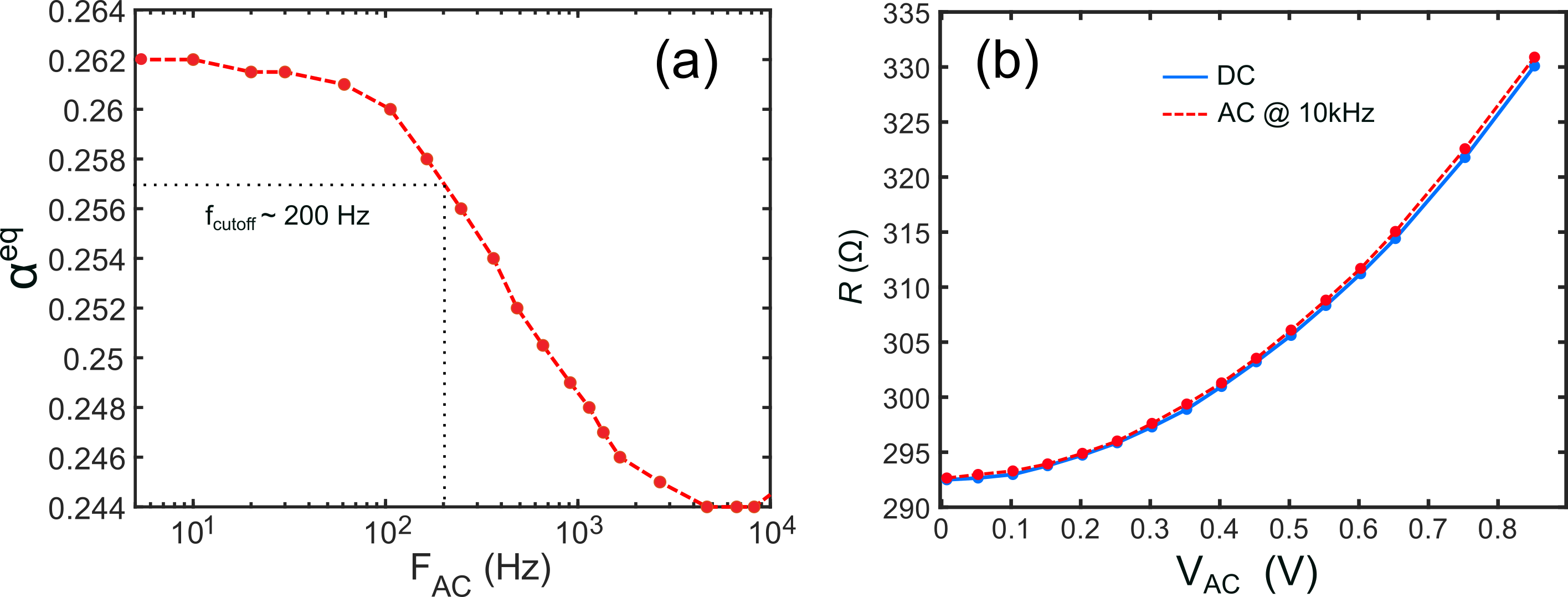}
    \caption{(a) Evolution of the balance parameter $\alpha^\mathrm{eq}$ as a function of the frequency. We observe a thermal cutoff at about \SI{200}{\hertz}. (b) Comparison between resistance measured with the Wheatstone bridge with a \SI{10}{\kilo \hertz} AC and a DC signal.} 
    \label{fig:AC_VS_DC}
\end{figure}

Fig.~\ref{fig:AC_VS_DC}(a) shows a typical evolution of $\alpha^\mathrm{eq}$ as a function of the frequency for a thermometer with 40 $\mu$m diameter glass sphere glued on it. It shows a thermal cutoff frequency at about \SI{200}{\hertz}. At large frequency ($f_\mathrm{AC}\simeq10$ kHz), the two additional components are strongly suppressed so that equations \ref{eq1}, \ref{eq2} and \ref{eq3} are valid. The resistance value can be deduced from $\alpha^\mathrm{eq}$ using equation \ref{eq2}. This is confirmed by Fig.~\ref {fig:AC_VS_DC}(b), which shows that the resistance values determined using equation \ref{eq2} operating the bridge either in DC or in AC at frequency $f_\mathrm{AC}\simeq10$ kHz are very close to each other.

\begin{figure}[!h]
    \centering
    \includegraphics[width=0.8\linewidth]{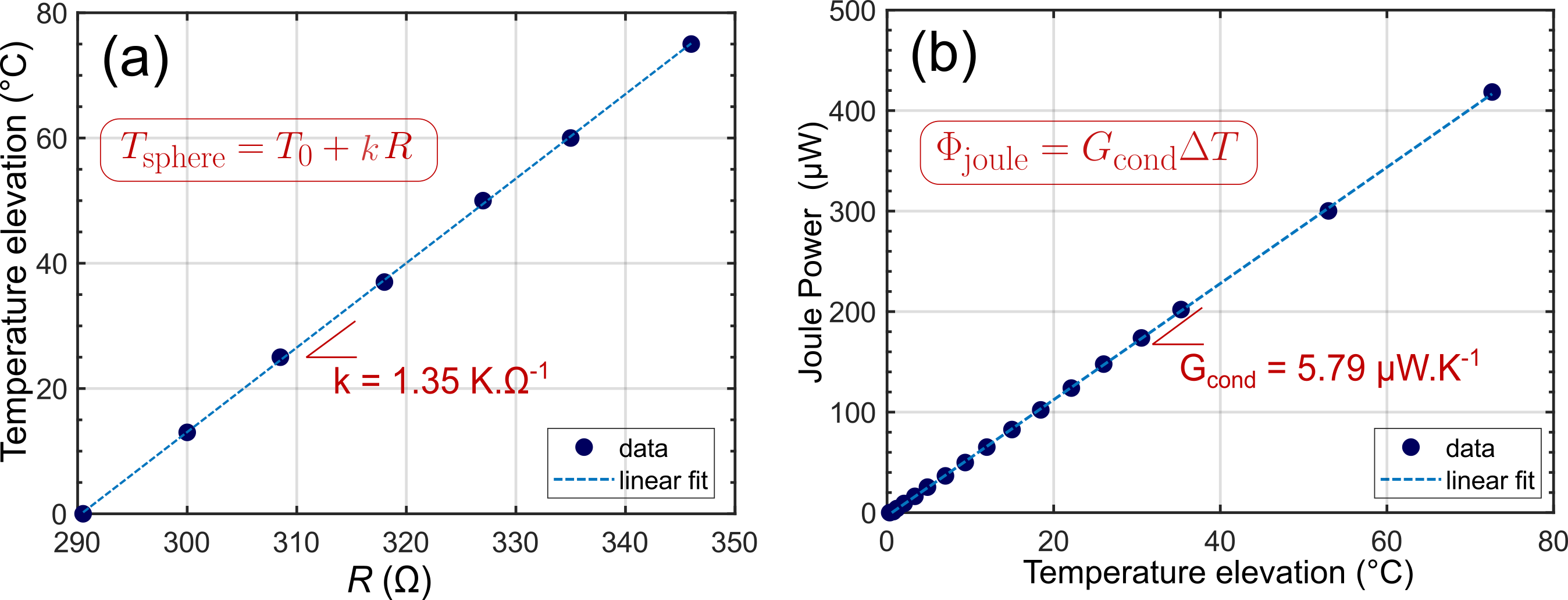}
    \caption{Thermal calibration of the probe. (a) Linear relation between the electrical resistance of the probe and the temperature elevation. (b) The relation between the Joule power and the temperature elevation gives the conductance through conduction of the cantilever.}
    \label{fig:calib}
\end{figure}

The probe has been calibrated to know the temperature, $T$, from the resistance value, $R$, of the palladium strip. To do so, we put the probe in a thermalized oven and we measure the resistance for each temperature. The resistance has been measured with an alternating AC current with a very low amplitude such that there is no Joule heating (equation \ref{eq2} is valid). In our temperature range, the relation between $T$ and $R$ is linear so we can extract the temperature coefficient from a linear fit (see Fig.~\ref{fig:calib}.(a)).

Once we know the relation between $R$ and $T$ we proceed to another measurement where we inject a current in the thermometer in vacuum and we measure the temperature elevation associated with this current. Because $Q_\mathrm{joule} = R I^2 $, and almost all of the flux goes through the cantilever we can deduce the conductance through the cantilever $G_\mathrm{cond}$ such that : 

\begin{equation}
    G_\mathrm{cond} = \frac{Q_\mathrm{joule}}{\Delta T} 
\end{equation}

\clearpage
\section{Measurements at different temperatures}

The same measurement of radiative heat transfer have been performed with different temperature differences with the planar substrate which is kept at ambient temperature ($T_0 = $ \SI{25}{\celsius}). The typical values of current injected in the probe to elevate the temperature of \SI{20}{\celsius}, \SI{40}{\celsius} and \SI{70}{\celsius} are respectively \SI{580}{\micro \ampere}, \SI{850}{\micro \ampere} and \SI{1100}{\micro \ampere}.

\begin{figure}[!h]
    \centering
    \includegraphics[width=\textwidth]{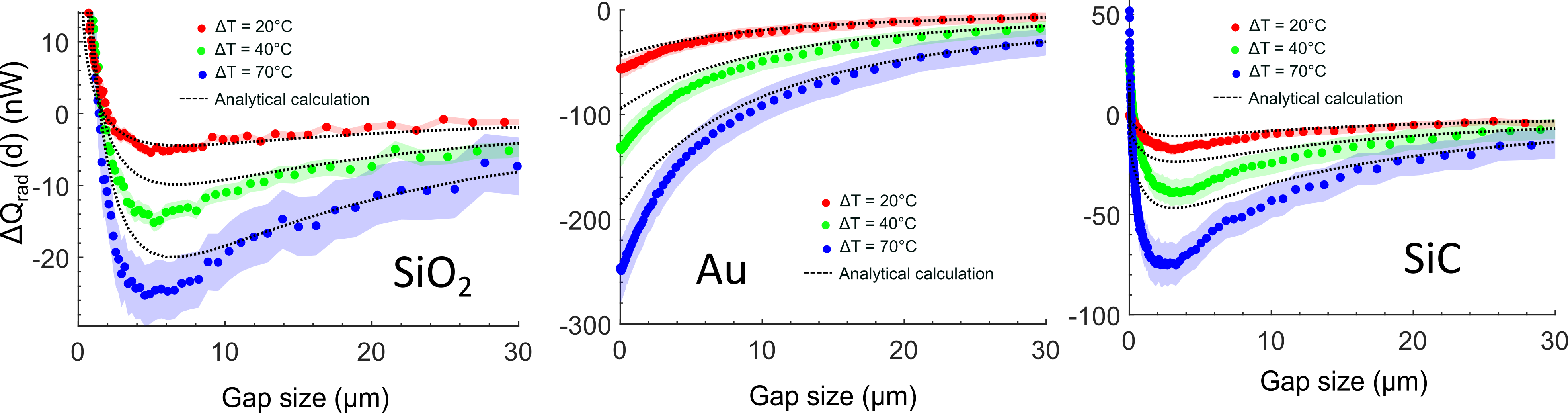}
    \caption{Measurement of the variation of radiative flux for three different temperatures of the microsphere. The dashed lines represent the theoretical calculation presented in the main text.}
    \label{fig:enter-label}
\end{figure}

We observe a good agreement, at least qualitative for the position and the amplitude of the flux variation for each set of temperatures and the three different materials. This agreement confirms the theoretical model developed in this paper.

\section{Radiative heat flux for the last micrometer}

In the manuscript, figure 3.a shows the clear enhancement of the flux for the silica substrate due to near field represented in a logarithmic scale. In the case of SiC substrate, the enhancement is much smaller, and in the case of gold we cannot even measure it. However, we show on the figure below, for the gold and SiC substrate the last 4 micrometers of flux measurement to see more clearly the effect or not of the near-field. Because the variation of flux $\Delta Q_\mathrm{rad} (d) $ has negative value, we represented the result in linear scale and not in logarithmic scale

\begin{figure}[!h]
    \centering
    \includegraphics[width=0.9\textwidth]{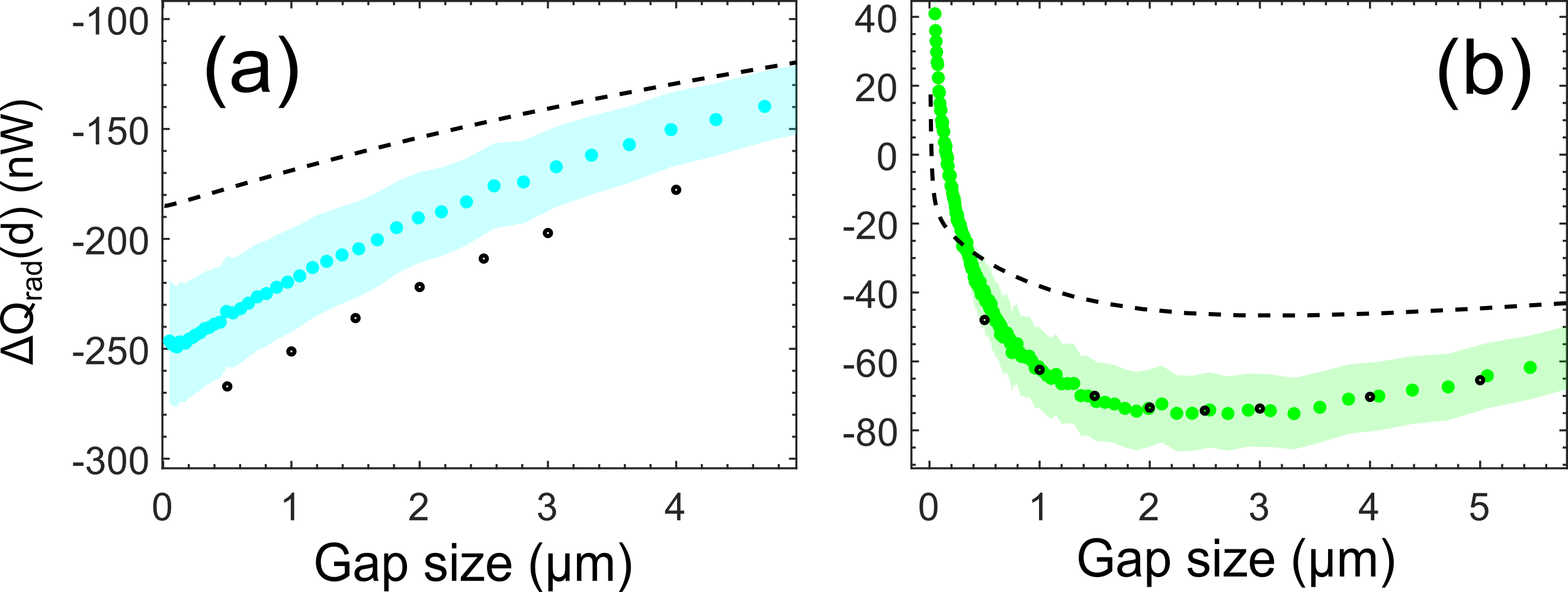}
    \caption{Variation of the radiative flux in the last micrometers before contact for the gold substrate (a) and the SiC substrate (b). The dashed line corresponds to the analytical model described in the manuscript. The Points are the calculation by SCUFF-EM}
    \label{fig:epsilons}
\end{figure}

\clearpage
\section{Dielectric functions and calculated substrate emissivities}

The dielectric functions that have been used in this paper for the three different materials are the following : 

\begin{figure}[!h]
    \centering
    \includegraphics[width=0.9\textwidth]{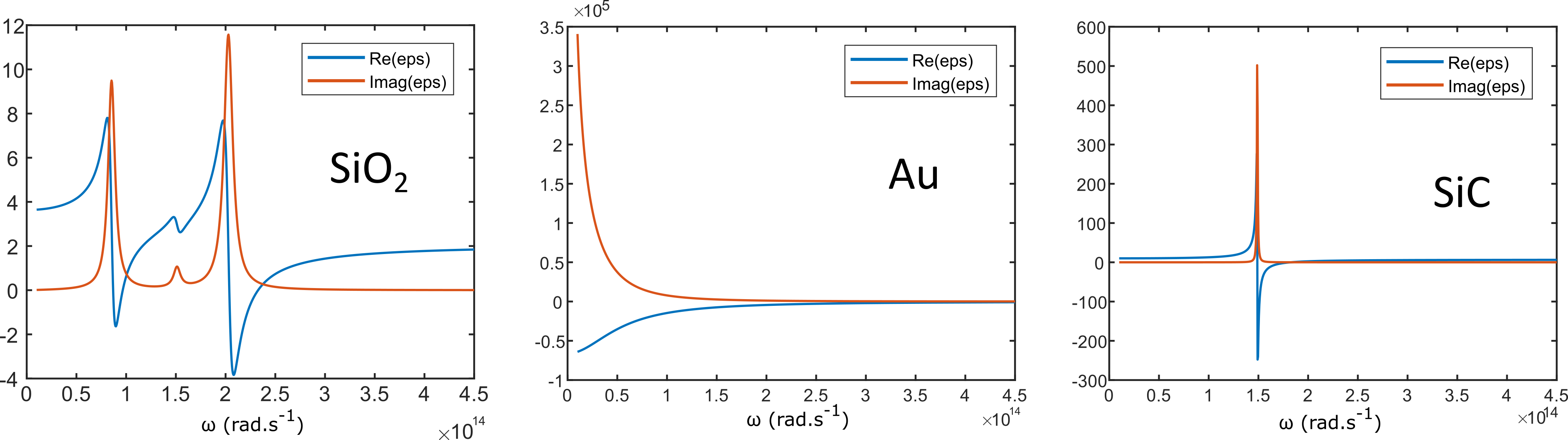}
    \caption{Real and imaginary part of the dielectric constants of the three materials used in this study.}
    \label{fig:epsilons}
\end{figure}

The far-field spectral emissivities of the planar substrates have been calculated according to the dielectric function : 

\begin{equation}
    \epsilon(z, \omega) = \int_{0}^{k_0} \kappa  \frac{(1 - \vert r^s_{12} \vert ^2) (1 - \vert r^p_{12} \vert ^2)}{2 k_0 \abs{k_\perp}}  \mathrm{d}\kappa  + \int_{k_0}^{ \infty} 4 \kappa^3  \frac{(1 -  \mathrm{Im}(r^s_{12}) ) (1 - \mathrm{Im}(r^p_{12}))}{2 k_0^3 \abs{k_\perp}} e^{-2 \mathrm{Im}(k_\perp) z} \mathrm{d}\kappa
\end{equation}

Where z is the distance with the interface. Here we choose $z = $ \SI{e-3}{\meter} to compute the far-field emissivity.

\begin{figure}[!h]
    \centering
    \includegraphics[width=0.9\textwidth]{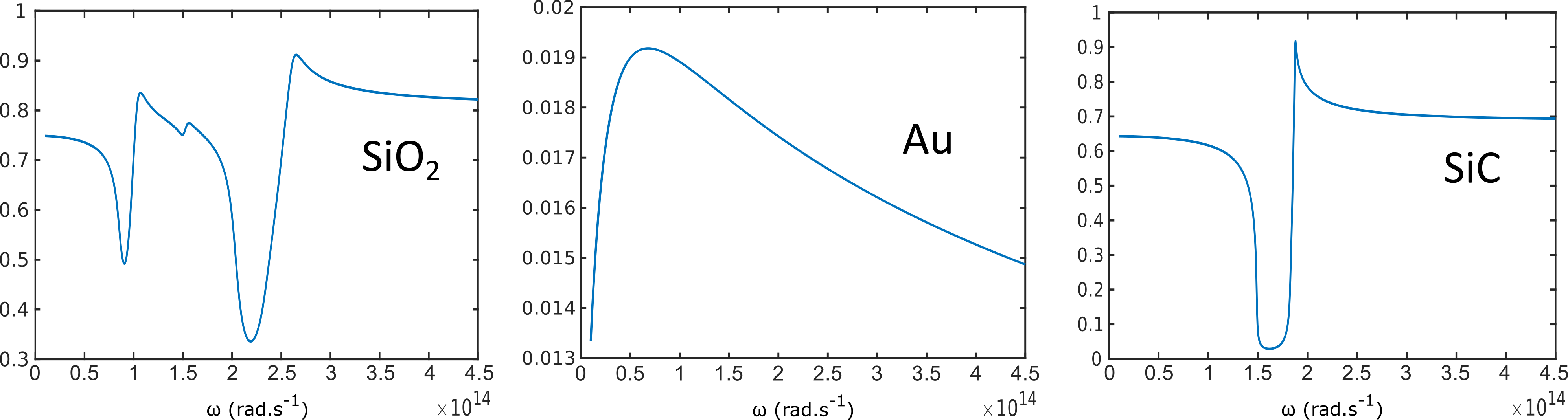}
    \caption{Calculated emissivities of the materials using the dielectric constants.}
    \label{fig:emissivities}
\end{figure}

\clearpage
\section{Emissivity of the silica microsphere}

The size of the sphere is on the order of the wavelength. Assimilating its emissivity to that of an infinite plane is not rigorous. The work of Kattawar [18], based on the formalism of fluctuational electromagnetism, allows for an exact calculation of the emission of a sphere with a dielectric function \(\varepsilon(\omega)\) and radius \(R\). It has notably been shown that, computing the emission of a sphere in thermodynamic equilibrium at temperature \(T\) is exactly equivalent to calculating the absorption efficiency \(Q_\mathrm{abs}\), as defined by Mie theory :

\begin{equation}
    \displaystyle Q_\mathrm{abs}={\frac {2}{x^2}}\sum _{n=1}^{\infty }(2n+1) \left[\Re (a_{n}+b_{n}) - \right |a_{n}|^{2}+|b_{n}|^{2} ]
\end{equation}
with, $x = k R$ is the size parameter often encountered in the Mie calculation. The coefficients $a_n$ and $b_n$ are defined as  : 
\begin{equation}
    a_n = \frac{\psi_n(x)}{\zeta_n(x)} \frac{D_n(y) - mD_n(x)}{D_n(y) - m G_n(x)}
\end{equation}
\begin{equation}
    b_n = \frac{\psi_n(x)}{\zeta_n(x)} \frac{mD_n(y) - D_n(x)}{mD_n(y) -  G_n(x)}
\end{equation}
where $y = m x$, $m$ the complex refractive index defined as $m= \sqrt{\varepsilon}$ and  $\psi_n(x)$, $\zeta_n(x)$, $D_n(y)$, $G_n(x)$ are the two Ricatti-Bessel functions and their logarithmic derivatives with respect to the argument of the function (see details in [18]). 

In figure S10.(a), we compare the emissivity of planar surface of silica and of a \SI{21.15}{\micro \meter} radius sphere. We see that between \SI{0.5}{\radian \cdot \s^{-1}} and \SI{3}{\radian \cdot \s^{-1}}, the calculation based on the infinite plan underestimate the radiative flux. For some frequencies, the emissivity is greater than one. This can be interpreted as an emission cross-section greater that the geometrical section of the sphere. Figure S10.(b) represents the calculated spectral flux emitted by a single sphere at \SI{370}{\kelvin} into an environment at $T_0 = $ \SI{300}{\kelvin}. We observe a strong agreement around \SI{99}{\percent} between the analytical calculation and the SCUFF-EM simulation. Conversely, we can see that taking the emissivity of a planar surface for the sphere underestimates the radiative flux of \SI{20}{\percent}.  

\begin{figure}[!h]
    \centering
    \includegraphics[width=0.9\linewidth]{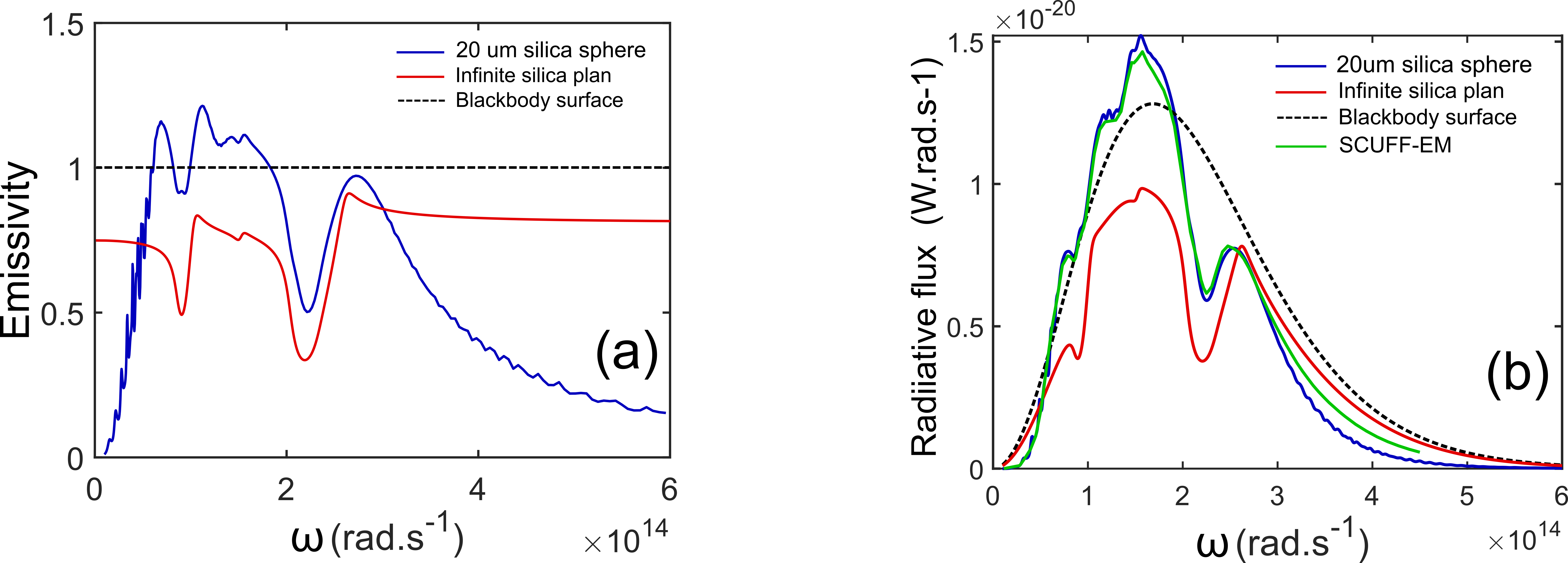}
    \caption{Comparison between thermal emission of an infinite planar surface and a \SI{21.15}{\micro \meter} radius silica sphere. (a) Emissivity (b) Radiative spectral flux}
    \label{fig:enter-label}
\end{figure}

\clearpage
\section{Exact heat transfer calculations for parallel surfaces}

Between two parallel surfaces, the conductance increases with decreasing gap size. Figure below represent the radiative heat transfer coefficient for parallel surface calculated in the framework of fluctuational electrodynamics. One surface is made of silica while the other is made of Gold, Silicone carbide or Silica. The dashed and pointed lines correspond respectively to evanescent and propagative waves. The colored one is the sum of the two contributions. One can notice that although the general behavior is the same for all three curves, the amplitude of variation is not. Contributions of surface modes between two silica surfaces is two orders of magnitude higher than if one is made of gold. In fact, gold’s surface polaritons are hardly excited at all at room temperature and they don’t match the ones of silica, the near-field contribution in the Au-SiO2 configuration is then very small and could be measurable only for nanometric gaps. Using our analytical model, we find a minimum of $\Delta Q_\mathrm{rad} (d)$ for the gold surface around $d= $ \SI{15}{\nano \meter}. Distance which is not achievable in practice. This explains why we only observe experimentally a decrease of the flux for the gold substrate.

\begin{figure}[!h]
    \centering
    \includegraphics[width=0.4\textwidth]{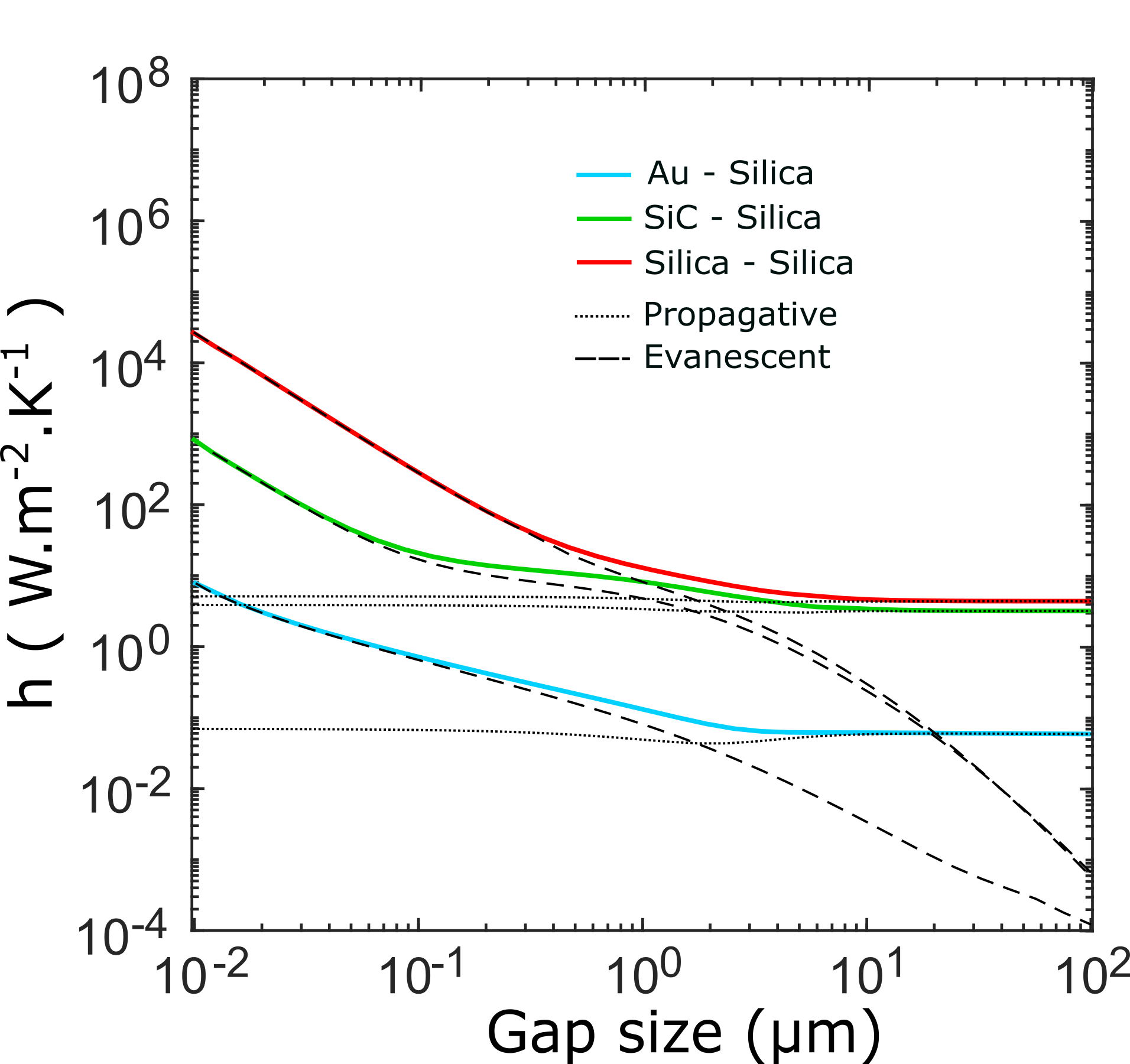}
    \caption{Radiative heat transfer coefficient for two parallel planes separated by a distance gap d calculated in the framework of fluctuational electrodynamics. The dashed and dotted lines are the contribution of evanescent and propagative waves.}
    \label{fig:parallel_surfaces}
\end{figure}

\clearpage
\section{Calculation using SCUFF-EM}

\begin{figure}[!h]
    \centering
    \includegraphics[width=0.89\textwidth]{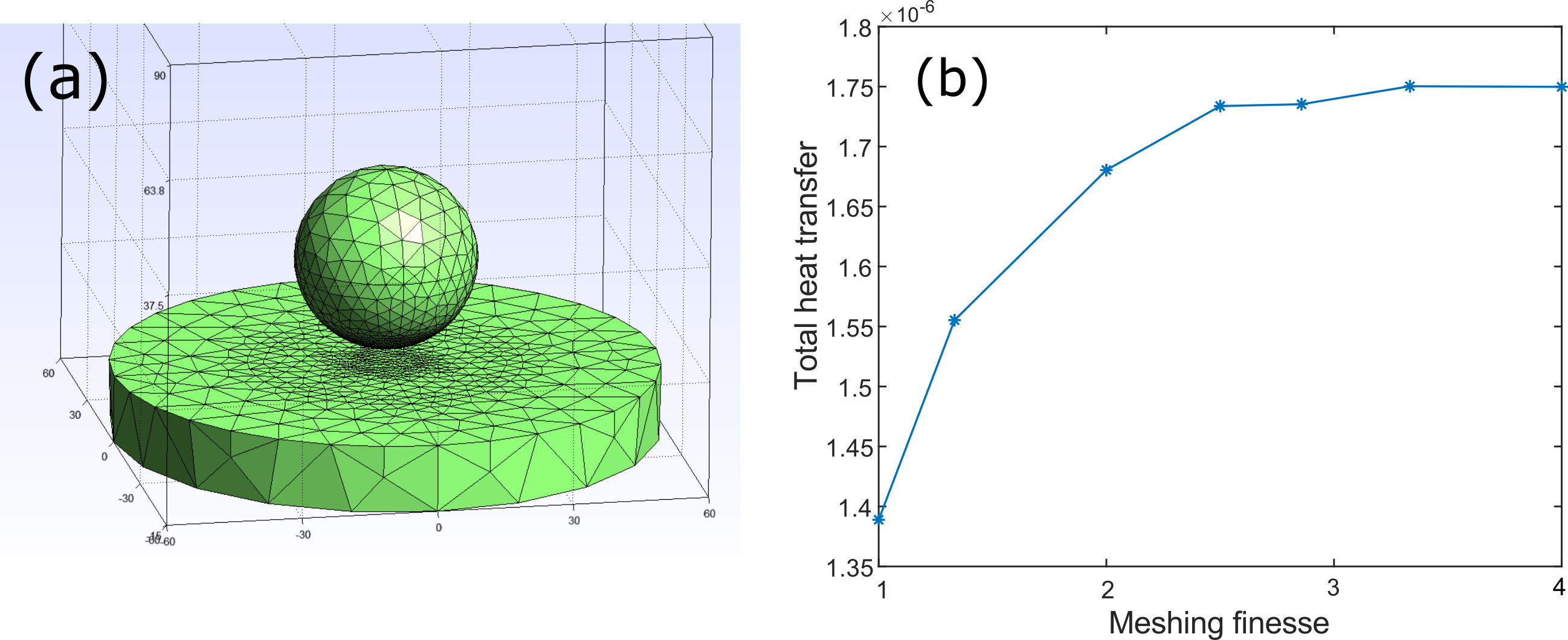}
    \caption{Geometry and meshing in SCUFF-EM simulations. (a). The optimal meshing used in our simulations. (b). Convergence test where a initial meshing is refined by a constant called the meshing finesse. We observe that the calculated flux converges after with a meshing finesse around 3.}
    \label{fig:scuff}
    \renewcommand{\figurename}{SUP}
\end{figure}

SCUFF-EM, a free open-source software implementation of the boundary-element method (BEM)
of electromagnetism scattering have been used to calculate the flux exchanged between the sphere and the plane. The geometry used is a \SI{40}{\micro \meter} diameter sphere in front of a large cylinder with a diameter of \SI{120}{\micro \meter} and a \SI{20}{\micro \meter} thickness (see Fig.~\ref{fig:scuff}.(a)). The meshing have been chosen to be finer at the bottom of the sphere and at the center of the cylinder where the distance between the two objects is minimal. We also choose a meshing where the bigger mesh element is smaller than the maximal wavelength. We proceeded to meshing convergence tests to be confident with the convergence of our results and to find the optimal meshing size. The results are shown on Fig.~\ref{fig:scuff}.(b) where the meshing finesse is the coefficient used to divide the mesh size. We see that the optimal meshing finesse is around 3, after that the computed heat transfer seems to tend toward a constant value.
Other calculations were performed with a \SI{240}{\micro \meter} diameter substrate to show that the non-monotonic behavior was fully captured with the \SI{120}{\micro \meter} diameter cylinder. The size of the substrate does not affect the radiative flux at short distances. However, a small difference of flux (few \SI{}{\nano \watt}) was observed when the sphere is at large distance to the substrate ($d > $ \SI{30}{\micro \meter}) for the silica substrate.

